\documentclass[twocolumn]{aa}
\usepackage{graphicx}
\usepackage{txfonts}

\def \arcsec   {\hbox{$^{\prime\prime}$}}   
\newcommand{\dfrac}[2]{\frac{\displaystyle{#1}}{\displaystyle{#2}}}

\newcommand{\stck}[1]{\stackrel{#1}{\longrightarrow}}

\begin{document}

\title{The physical and chemical structure of hot molecular cores}

\author{H. Nomura \and T. J. Millar}
\institute{Department of Physics, UMIST, PO Box 88, 
Manchester M60 1QD, UK \\
\email{h.nomura@umist.ac.uk, tom.millar@umist.ac.uk}}

\date{Received 21 Jyly 2003/ Accepted 21 October 2003}

\abstract{
We have made self-consistent models of the density and temperature
profiles of the gas and dust surrounding embedded luminous objects
using a detailed radiative transfer model together with observations of
the spectral energy distribution of hot molecular cores. Using these
profiles we have investigated the hot core chemistry which results
when grain mantles are evaporated, taking into account the different
binding energies of the mantle molecules, as well a model  
in which we assume that all molecules are 
embedded in water ice and have a common binding energy. We find that 
most of the resulting
column densities are consistent with those observed toward
the hot core G34.3+0.15 at a time around 10$^4$ years after central
luminous star formation. We have also investigated the dependence of
the chemical structure on the density profile which suggests an
observational possibility of constraining density profiles from
determination of the source sizes of line emission from desorbed
molecules.
}

\maketitle
%

\section{Introduction}

Hot cores are compact (diameters $\leq 0.1$ pc), dense (n$\geq 10^7$
cm$^{-3}$), hot (T$\geq 100$ K), and dark ($A_v\geq 100$ mag) molecular
cloud cores. At least some of them are believed to be precursors to
ultracompact HII regions and in the early stage of forming massive stars
because luminous infrared sources, massive bipolar outflows
and/or inflow motion are observed associated with them (e.g., Henning
et al. 2000; Kurtz et al. 2000; Churchwell 2002). They are
observationally characterized by anomalously large abundances of some
molecular species including hydrogenated, saturated, small molecules
such as H$_2$O and H$_2$S, and large complex molecules such as
HCOOCH$_3$ and (CH$_3$)$_2$O, in contrast to colder dark clouds. It is
thought that these molecules are produced under conditions specific
to star formation: the freeze-out of gas-phase species onto
dust grains and the grain surface reactions in the cold collapsing
pre-protostellar phase, and the evaporation of the grain mantle material 
and the subsequent gas-phase reactions in the hot protostellar phase
(e.g., Millar 1993, 1997). Therefore, the observed large column
densities of some molecules in hot cores are also expected to be an
indicator of luminous high mass star formation which yields a large
mass of high temperature gas. Previously, chemical models of hot cores
have been studied together with the pre-stellar phase or
with some initial conditions of evaporated grain mantle species, and
have been successful in reproducing their observational properties (e.g., Brown
et al. 1988; Millar et al. 1991; Charnley et al. 1992; Caselli et
al. 1993; Charnley 1997; Millar et al. 1997; Viti \& Williams 1999;
Rodgers \& Charnley 2001; Doty et al. 2002).

Physical properties of hot cores have been investigated on the basis of
observational studies of molecular line and dust continuum emission.
Some physical models together with radiative transfer calculation and/or
chemical calculation have succeeded in reproducing the observations of
spectral energy distributions and/or molecular column densities (e.g.,
Millar et al. 1997; Kaufman et al. 1998; Osorio et al. 1999; Hatchell et
al. 2000; van der Tak et al. 2000a; Doty et al. 2002).
On the other hand, the dynamical processes and the formation mechanisms of
massive stars are not yet established because of the difficulty of
treating in a self-consistent manner young stellar evolution, the accretion 
flow
and radiation pressure. Non-spherical accretion to protostellar cores,
for example, through accretion disks, and/or coalescence of lower mass
stars may be involved in massive star formation, but there are, as yet, no
accepted theories (e.g., Bonnell et al. 1998; Stahler et al. 2000; Yorke
\& Sonnhalter 2002). The nature and timescale of the dissipation of the
surrounding envelopes of young massive stars, for example, by powerful
outflows and ionizing strong winds which stop the accretion, are also
uncertain but important in determining the mass of forming
stars in the case of the isolated star formation (e.g., Shu et al. 1987;
Nakano et al. 1995).

In this paper, we have modeled the density and temperature profiles of
hot cores, making use of radiative transfer calculations, and then
investigated the hot core chemistry, taking into account the temperature
dependent grain mantle evaporation. In the following section, we present the
physical model of the density profiles constrained by observational results
and the temperature profiles calculated by solving the radiative transfer
equation. In Sect. 3 we introduce the chemical model along with the
temperature dependent grain mantle evaporation process. These models are
applied to the hot core G34.3+0.15, and the results of chemical
calculation are described and compared with the observations in Sect. 4.
Also, we investigate the dependence of the chemical structure of hot
cores on (1) the trapping of mantle molecules in water ice and (2)
the density profile generally to discuss the chemical and physical condition
of massive-star-forming clumps in Sect. 5. Finally, the results are summarized
in Sect. 6.

\section{Physical model}

In this paper we consider a spherically symmetric molecular 
clump illuminated by 
a central star with luminosity $L_*$. The density profiles modeled are
constrained by observational results (Sect. 2.1), and the temperature 
profiles are calculated self-consistently under the assumption of 
radiative equilibrium by solving the radiative transfer equation
(Sect. 2.2). 

\subsection{Density profiles}

As we will see in Sect. 5, the density profile of a clump is one of
the crucial properties which affects its chemical structure. However,
there is no established theoretical model because the dynamical process
of massive star formation is still unknown
due to the difficulty in comprehensive treatments of 
young stellar evolution and accreting flow as well as ionizing
radiation, radiation pressure, stellar wind, bipolar outflow, and so
forth (e.g., Stahler et al. 2000; Yorke \& Sonnhalter 2002). 
Thus, in this paper we model the density profiles of clumps simply based
on the submillimetre observations by Hatchell et al. (2000), avoiding
uncertainties in theoretical models.

Hatchell et al. (2000) investigated spectral energy distributions (SEDs)
and radial profiles of 450 and 850 $\mu$m continuum dust emission of five
ultracompact HII regions to conclude that chemically rich hot molecular
clumps can be modeled by a combination of compact high density cores
plus $r^{-1.5}$ density profile envelopes. 
Also, van der Tak et al. (2000a) studied 14 envelopes around young massive
stars by means of CO, CS, and H$_2$CO line emission maps, and SEDs and
radial profiles of submm dust emission to find that the best-fit values
of the power law indices for density profiles of the envelopes
range from $-1.0$ to $-1.5$.
Theoretically, an isothermal self-gravitating gas is known to have
$r^{-2}$ density profile via the balance between self-gravity and pressure
gradient forces (e.g., Ebert 1955; Bonnor 1956; Larson 1969; Shu 1977).
If the nonthermal velocity dispersion, which has been observed in line
widths towards various sizes of interstellar clouds and known to
have a scaling-law of $\delta v\propto r^{0.5}$ (e.g., Larson 1981;
Falgarone et al. 1992; Caselli \& Myers 1995), affects the
pressure gradient force, it may flatten the radial density profiles of
clumps like the logotrope model (e.g., McLaughlin \& Pudritz 1996). 

Taking into account of these observational and theoretical studies,
we use the following spherically symmetric hydrogen number density
distribution in this paper: 
\begin{equation}
n(r)=\left\{\begin{array}{ll}
n_0/(1+r/r_{\rm c})^2+n_1/(1+r/r_{\rm tr})^{1.5} & {\rm if}\ r_{\rm c}<r_{\rm tr},  \\
n_1/(1+r/r_{\rm c})^{1.5} & {\rm otherwise},
\end{array}\right. \label{eq1}
\end{equation}
where the radius of $r_{\rm tr}=0.05$ pc is adopted for the transition
between isothermal and nonthermal regions, according to the observations
of molecular line widths (e.g., Fuller \& Myers 1992; Myers \& Fuller
1992). The hydrogen column
density of the clumps, $N_{\rm H}\equiv 2\int_{R_*}^{r_{\rm
out}}n(r)dr$, is fixed as $N_{\rm H}=10^{25}$ cm$^{-2}$, based on the 
observations of the chemically rich hot molecular clumps by Hatchell et
al. (2000). The inner radius is set at the stellar
surface, $R_*$, and the outer radius is set at $r_{\rm out}=1.0$ pc. The
central densities $n_0$ and $n_1$ are determined to satisfy the relation
of $n_0/(1+r_{\rm tr}/r_{\rm c})^2=n_1/(1+r_{\rm tr}/r_{\rm tr})^{1.5}$
and the column density of $N_{\rm H}=10^{25}$ cm$^{-2}$.

\subsection{Temperature profiles}

The temperature profiles are calculated self-consistently, assuming that
the heating source of the clump is a central star with
luminosity $L_*$, under the condition of local radiative
equilibrium, that is, the balance between absorption and
reemission of radiation by dust grains at each point in the clump:
\begin{equation}
4\pi{\displaystyle\int_0^{\infty} d\nu \kappa_{\nu}B_{\nu}[T(r)]
=\int_0^{\infty} d\nu \kappa_{\nu}
\oint d\mu d\phi I_{\nu}(r;\mu,\phi)},  \label{eq2}
\end{equation}
where $T$, $\kappa_{\nu}$, and $B_{\nu}(T)$ represent the temperature,
the monochromatic opacity, and the Planck function for blackbody
radiation at a frequency $\nu$, respectively. Local thermodynamic
equilibrium, $\eta_{\nu}=\kappa_{\nu} B_{\nu}(T)$, is adopted, where
$\eta_{\nu}$ is the monochromatic emissivity. The specific intensity
$I_{\nu}$ is calculated by solving the radiative transfer equation, 
\begin{equation}
I_{\nu}(r;\mu,\phi)=\int_0^s \kappa_{\nu}(r')\rho(r')B_{\nu}[T(r')]e^{-\tau_{\nu}(r')}ds', 
\end{equation}
where $\tau_{\nu}(r')$ is the specific optical depth from $r'$ to
$r$. The short characteristic method in spherical coordinates (Dullemond
\& Turolla 2000) is used to compute the integration.
The radiative transfer equation is solved only in the dusty region
(where the dust grains are not thermally destroyed) in this calculation.
The inner radius for the numerical calculation is determined as
$r_{\rm in}=(T_{\rm in}/T_*)^{-2}R_*$, where $T_*$ and $R_*$ are the
temperature and radius of the central star, assuming that the radial
profile of the temperature in the dust-free region ($r<r_{\rm in}$) is
given by $T(r)\propto r^{-1/2}$, which represents the profile in the
optically thin region with gray opacity heated by a point source.
For the dust opacity model, we adopt the frequency dependent absorption 
coefficient $\kappa_{\nu}$ of Adams \& Shu (1985, 1986), which
approximates the interstellar dust grains model by Draine \& Lee
(1984). The dust destruction temperature is set as $T_{\rm in}=$ 2300K
in this model (see Nomura 2002 for the details of radiative transfer
calculation). In this paper the gas temperature is assumed to be the
same as the dust temperature, based on the result of our calculation that
the difference between the gas and the dust temperatures is small, 
except for the very inner region of the clump, where the gas
temperature is $\geq 10^3$ K due to photoelectric heating from small
grains induced by the ultraviolet radiation from the central star
(cf. Doty \& Neufeld 1997).

\section{Chemical model}

It is generally believed that the relatively large abundances of 
hydrogenated, saturated, small molecular
species, such as H$_2$O and H$_2$S, and large complex molecules, such as 
HCOOCH$_3$ and (CH$_3$)$_2$O, observed in hot cores are
related to the evolution of young stellar objects and the evaporation
of grain mantle material. 
In the cold, pre-stellar phase of the dense molecular clumps
the accretion timescale of gas-phase atoms and molecules onto grain
surface, $\sim 3\times 10^9/n$ [cm$^{-3}$] yr at $T=$10 K, is short
enough since the density, $n$, is high via self-gravitating
contraction of the clumps. Thus, the accretion of gas-phase species
onto grain surfaces and subsequent surface chemistry produce grain
mantles rich in saturated species. Once massive star formation
occurs in the clumps, the molecules are evaporated from the grain
surface and the subsequent gas-phase reactions produce the large complex
molecules seen in the hot gas
(see e.g., reviews by Millar 1993, 1997 and references therein), 
although there is some evidence that even the most complex of molecules
may have been formed on grains.
On the other hand, the recent detailed analyses of observations of ice
absorption features in infrared spectra towards massive young stellar
objects have suggested that icy grain mantles in envelopes surrounding
the central stars have different composition at different temperatures
along the line-of-sight, that is, each icy mantle molecule sublimates
from dust grains at a specific temperature (e.g., Ehrenfreund et
al. 1998). 
In this paper, we set the molecular species evaporated from grain
mantles as an initial condition, and then calculate the subsequent
time-dependent gas-phase chemical reactions, 
taking into account the different binding energies of the mantle
molecules. 

The initial fractional abundances of species with respect to total
hydrogen nuclei listed in Table 1 are used in this paper.
The initial gas is assumed to be mostly molecular and slightly ionized
by cosmic rays. The initial mantle composition is
determined so as to be consistent with the recent observations of ice
absorption features in infrared spectra towards massive young stellar
objects (Gerakines et al. 1999; Keane et al. 2001)
and other hot core models (Charnley et al. 1992; Millar et al. 1997;
Rodgers \& Charnley 2001; Doty et al. 2002), and fixed so as to best fit
the results of model calculations to observations towards
G34.3+0.15 (see Sect. 4).
Also, the total abundances of carbon, oxygen, and nitrogen 
are set to agree with the mean interstellar values of C/H$=1.4\times
10^{-4}$, O/H$=3.19\times 10^{-4}$, and N/H$=7.5\times 10^{-5}$
(Cardelli et al. 1996; Meyer et al. 1997; Meyer et al. 1998).

\begin{table}
\caption[]{Initial abundances and injection radii of mantle molecules in
 the model of G34.3+0.15.}\label{T1}
$$ \begin{array}{p{0.11\linewidth}cc|p{0.12\linewidth}cc} \hline 
 Species & {\rm Abundance} & r_{{\rm inj},i} [{\rm pc}] & Species & {\rm Abundance} & r_{{\rm inj},i} [{\rm pc}] \\ \hline
 H$^+$ & 1.0\ 10^{-11} & & CO & 1.3\ 10^{-4} & \\
 He$^+$ & 2.5\ 10^{-12} & & CO$_2$ & 3.0\ 10^{-6} & 0.18 \\
 H$_3^+$ & 1.0\ 10^{-9} & & H$_2$CO & 2.0\ 10^{-6} & 0.3 \\
 Fe$^+$ & 2.4\ 10^{-8} & & CH$_3$OH & 2.0\ 10^{-7} & 0.12 \\
 He & 0.1 & & C$_2$H$_5$OH & 5.0\ 10^{-9} & 0.14 \\
 S & 5.0\ 10^{-9} & & O$_2$ & 1.0\ 10^{-6} & 0.5 \\
 Si & 3.6\ 10^{-8} & & H$_2$O & 2.8\ 10^{-4} & 0.08 \\
 C$_2$H$_2$ & 5.0\ 10^{-7} & 0.22 & N$_2$ & 3.7\ 10^{-5} &  \\
 CH$_4$ & 2.0\ 10^{-7} & 0.7 & NH$_3$ & 6.0\ 10^{-7} & 0.17 \\
 C$_2$H$_4$ & 5.0\ 10^{-9} & 0.21 & H$_2$S & 1.0\ 10^{-7} & 0.29 \\
 C$_2$H$_6$ & 5.0\ 10^{-9} & 0.20 & OCS & 5.0\ 10^{-8} & 0.17 \\ \hline
\end{array}
$$ 
\end{table}

Each surface species, $i$, is expected to evaporate thermally with
the timescale of 
\begin{equation}
\tau_{{\rm evap},i}=\nu_{0,i}^{-1}\exp(E_{b,i}/kT), \label{eq4}
\end{equation}
where $k$ and $T$ are the Boltzmann constant and the temperature,
respectively. 
The vibrational frequency $\nu_{0,i}$ is given by $\nu_{0,i}=(2n_s
E_{b,i}/\pi^2m_i)^{1/2}$, where $n_s\approx (7{\rm \AA})^{-1}$ is the
number of sites per unit surface area and $m_i$ the mass of the species
(e.g., Tielens \& Allamandola 1987). The binding energies $E_{b,i}$ are 
taken from Hasegawa \& Herbst (1993),  Aikawa et al. (1997), Willacy et
al. (1998), and Fraser et al. (2001) (see also references therein). 
Now, the temperature in the clump is a decreasing function of distance
from the central star and assumed to be unchanged with time, while
the evaporation rate is very sensitive to the temperature as can be seen
from Eq. (\ref{eq4}). For example, in the case of water, whose
binding energy is
$E_{b,{\rm H}_2{\rm O}}/k=5773$ K, a temperature rise from $T=95$K to 105K
shortens the evaporation time by about 1/100. Thus, in this paper the
evaporation process is simply modeled:- each species, $i$, evaporates
instantaneously inside a radius $r_{{\rm inj},i}$ where the evaporation
time $\tau_{{\rm evap},i}$ is shorter than the accretion time, 
\begin{equation}
\tau_{{\rm acc},i}=(S\pi a^2 d_gnv_i)^{-1}, \label{5}
\end{equation}
while it is retained on the dust otherwise.
That is, the gas-phase abundances of mantle molecules are initially
set as those listed in Table 1 inside the injection radii $r_{{\rm
inj},i}$, and zero outside these radii. Beyond a radius of 1 pc, the
density becomes so low that the accretion time becomes comparable to or
greater than the chemical time-scales so that mantle formation is not 
efficient. We do not model this extended envelope here.
In Eq. (\ref{5}) $n$, $a$, $d_g$, and $v_i=(2kT/m_i)^{1/2}$ are
the number density of hydrogen nuclei, the grain radius, the ratio of
the number density of grains to $n$, and
the thermal velocity of the species, $i$, respectively. The
sticking probability $S$ is set as $S=0.3$. For the grain surface area
per hydrogen nucleon, $\pi d_ga^2$, we used the value of
$<d_ga^2>=2.2\times 10^{-22}$cm$^2$ (see Rawlings et al. 1992 for
the parameter values).
The corresponding injection radii $r_{{\rm inj},i}$ in the model of
G34.3+0.15 (see next section) are also listed in Table 1.
The binding energies of CO and N$_2$ are low enough and the kinetic
temperature high enough that they
exist in the gas-phase throughout the region modelled here. 

The subsequent gas-phase chemistry is calculated time-dependently by
means of the chemical network based on that of Millar et al. (1997),
which consists of 209 species connected by 2191 reactions. The standard 
interstellar cosmic-ray ionization rate, $\zeta=1.3\times
10^{-17}$s$^{-1}$, is adopted. The
adsorption process of neutral atoms and molecules $i$ onto grain surface
is included only if $\tau_{{\rm acc},i}<\tau_{{\rm evap},i}$ in this
calculation. The accretion rate is given by
\begin{equation}
k_{{\rm acc},i}=-S\pi a^2 d_gnv_in_i,
\end{equation}
where $n_i$ is the number density of species $i$. This process
affects the gas-phase abundances slightly in the outer region, but
is not very significant because the accretion rate of each
species is comparable to the chemical reaction rates.

\section{Model of hot core G34.3+0.15}

In this section, we model the physical and chemical structure of the hot
molecular core G34.3+0.15 using the radiative transfer and the
pseudo-time-dependent gas-phase chemistry calculations described in the
previous sections together with observations of the spectral energy
distribution and the molecular line emissions toward G34.3+0.15. The hot
core associated with this source consists of a small clump of hot gas 
situated at the nose of a cometary-shaped, ultra-compact H{\sc ii} region.
It has been well-observed, particularly through spectral line surveys
(Macdonald et al. 1996; Kim et al. 2000) at 1, 2 and 3 mm, and has a rich 
molecular line spectrum, with large abundances of complex, highly-saturated
molecules.

\subsection{Density and temperature profiles}

\begin{figure}
\centering
\resizebox{\hsize}{!}{\includegraphics{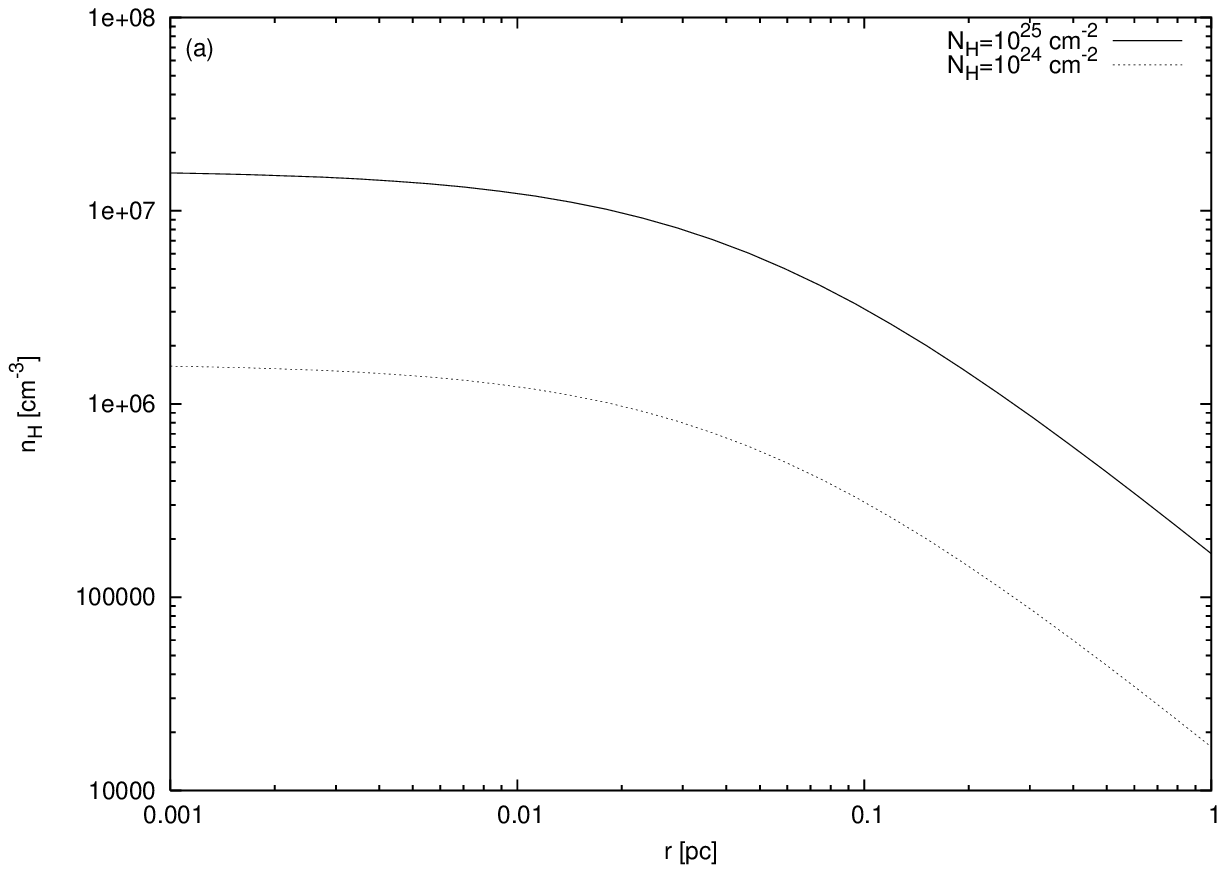}}
\resizebox{\hsize}{!}{\includegraphics{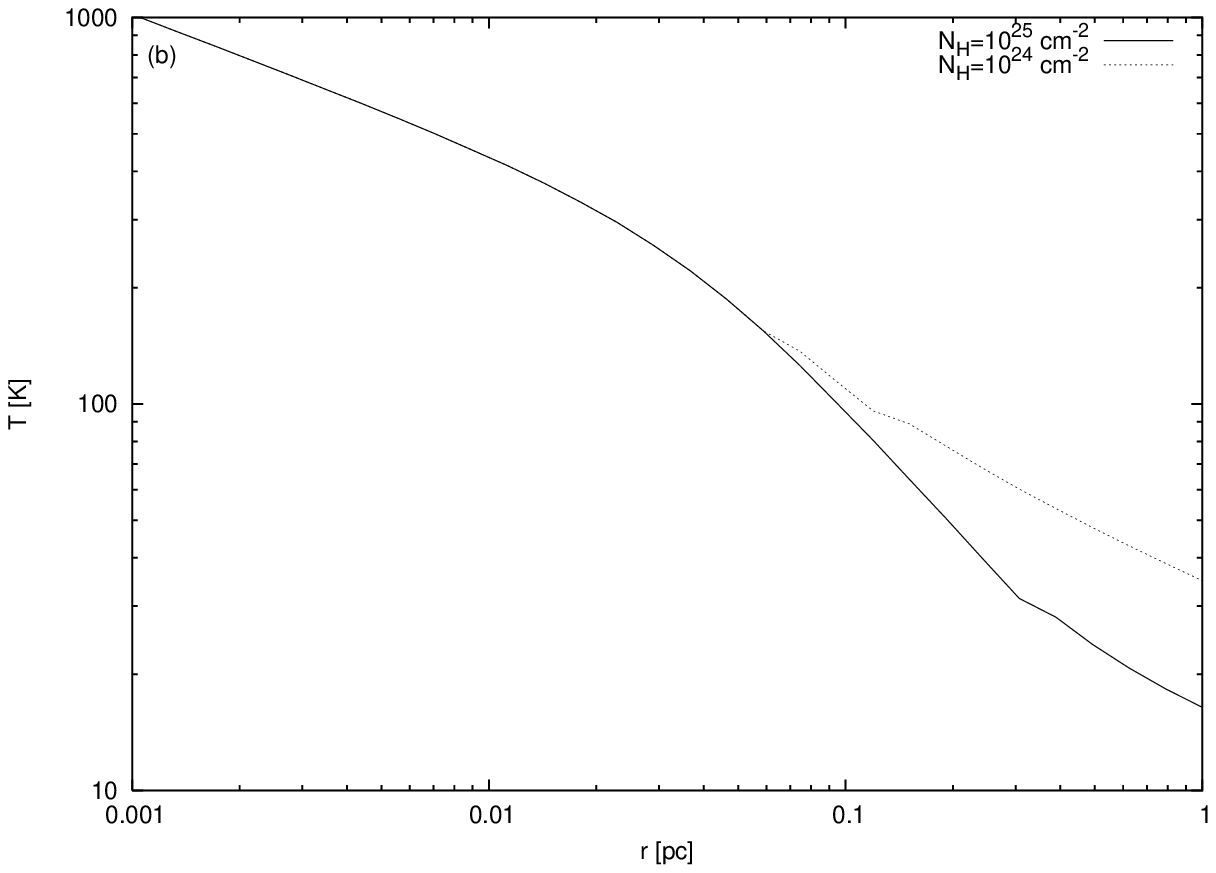}}
\resizebox{\hsize}{!}{\includegraphics{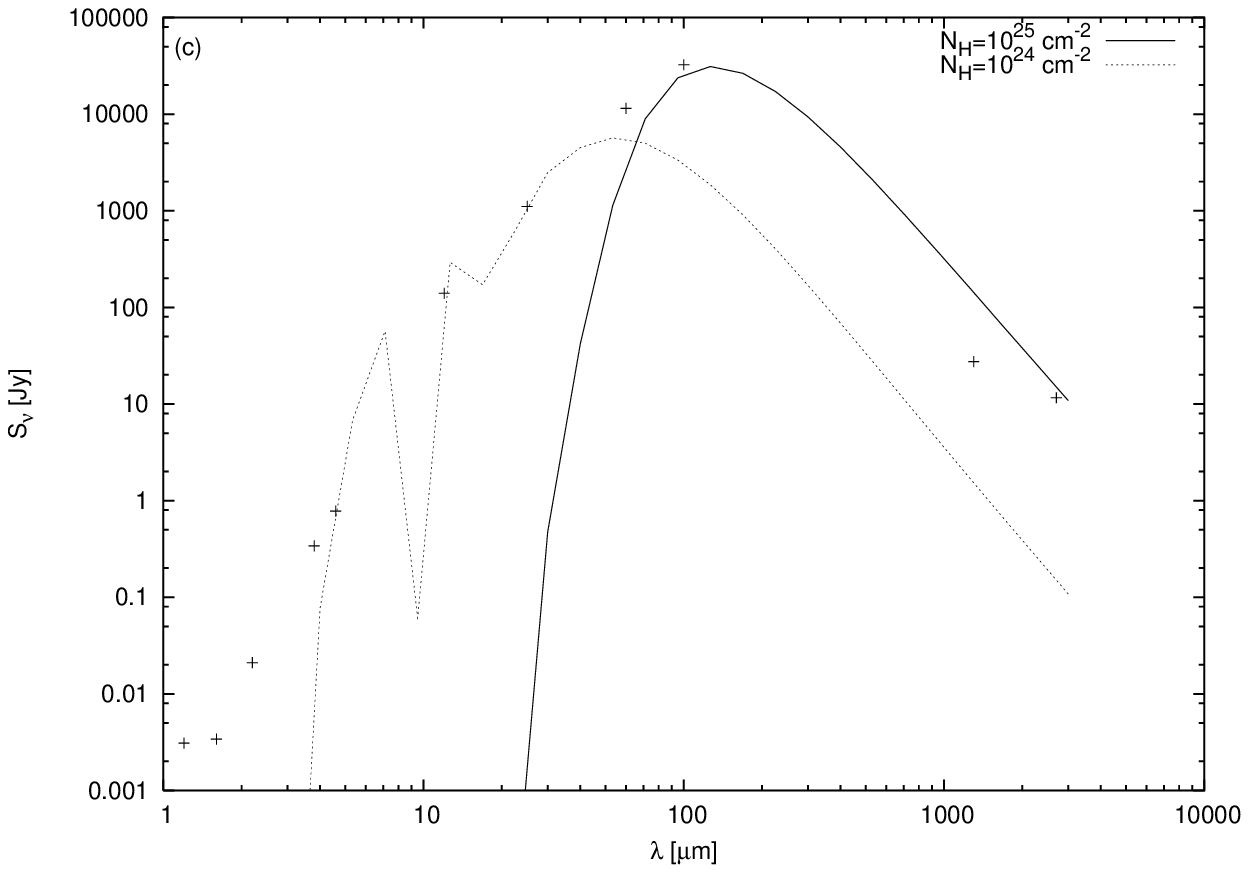}}
\caption{(a) Density and (b) temperature profiles and (c) SED of the
 model of G34.3+0.15. Combination of a envelope with $N_{\rm H}=10^{25}$
 cm$^{-2}$ (solid lines) and a cavity with $N_{\rm H}=10^{24}$ cm$^{-2}$
 (dotted lines), whose core radii are $r_{\rm c}=0.05$ pc,
 reproduces the SED observational data (crosses).}
\end{figure}

The density and temperature profiles of the hot core G34.3+0.15 are
modeled so that the theoretical calculation of the spectral energy
distribution (SED) is consistent with that observed. We take $r_{\rm
c}$ defined in Sect. 2.1 as a parameter and obtain its best-fit value by
iteratively calculating the temperature profile (see Sect. 2.2) and
comparing the resulting SED with observation. The luminosity of the
central star is set as $L_*=4.7\times 10^5 L_{\odot}$, referring to
Watt \& Mundy (1999). For the temperature and radius of
the star we adopt $T_*=4.5\times 10^4$K and $R_*=11R_{\odot}$, 
assuming a massive zero-age main-sequence star (e.g., Thompson 1984; see
also references therein). The distance to G34.3+0.15 of 3.7 kpc and the
observational data for the SED are taken from Chini et al. (1987).
As a result of the calculations, we find that the model with $r_{\rm
c}=0.05$ pc represents the best fit to the observational data at 
wavelengths from
far-infrared to millimeter. The resulting density and temperature
profiles, and the calculated SED are shown in Figs.1a, 1b, and 1c,
respectively, with solid lines. The crosses in Fig.1c are the
observational data. 

\begin{figure}
\centering
\resizebox{\hsize}{!}{\includegraphics{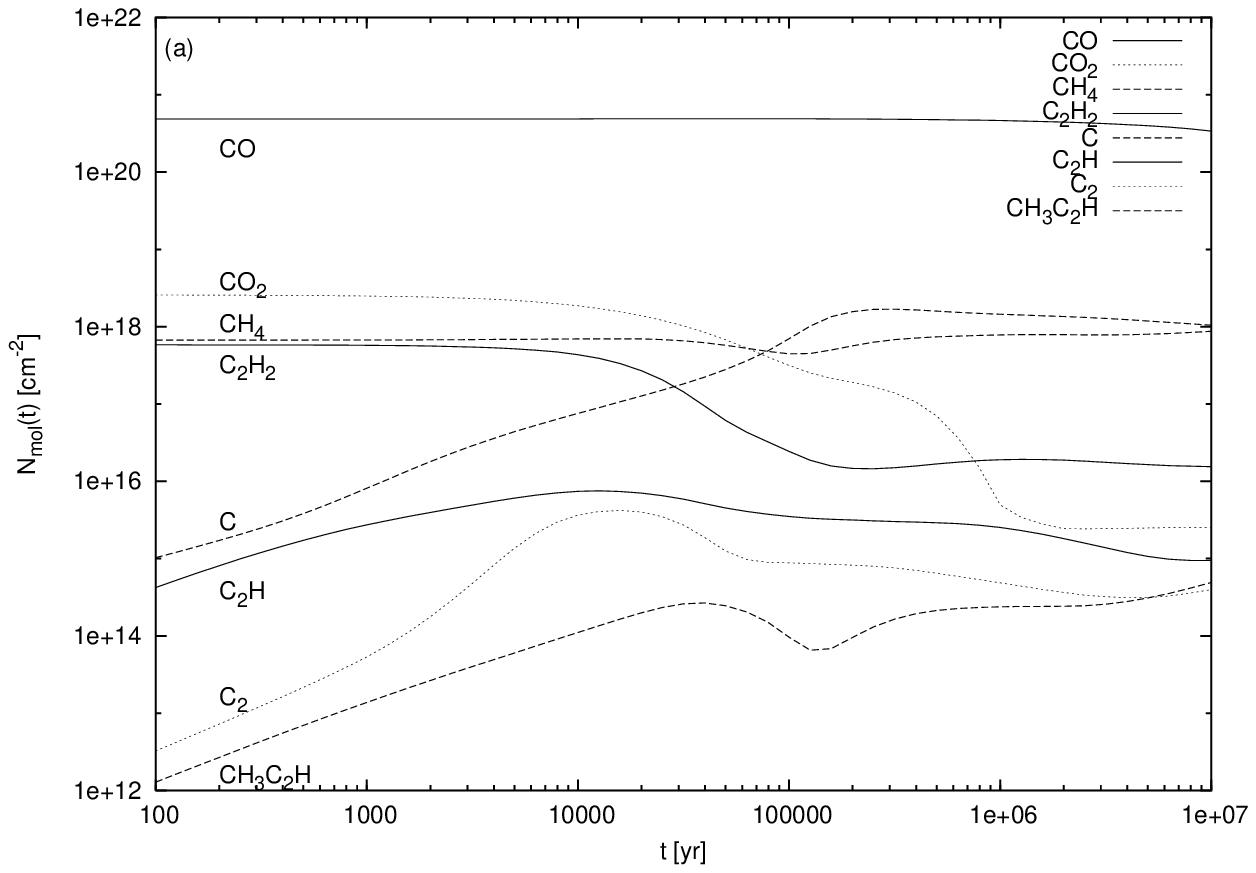}}
\resizebox{\hsize}{!}{\includegraphics{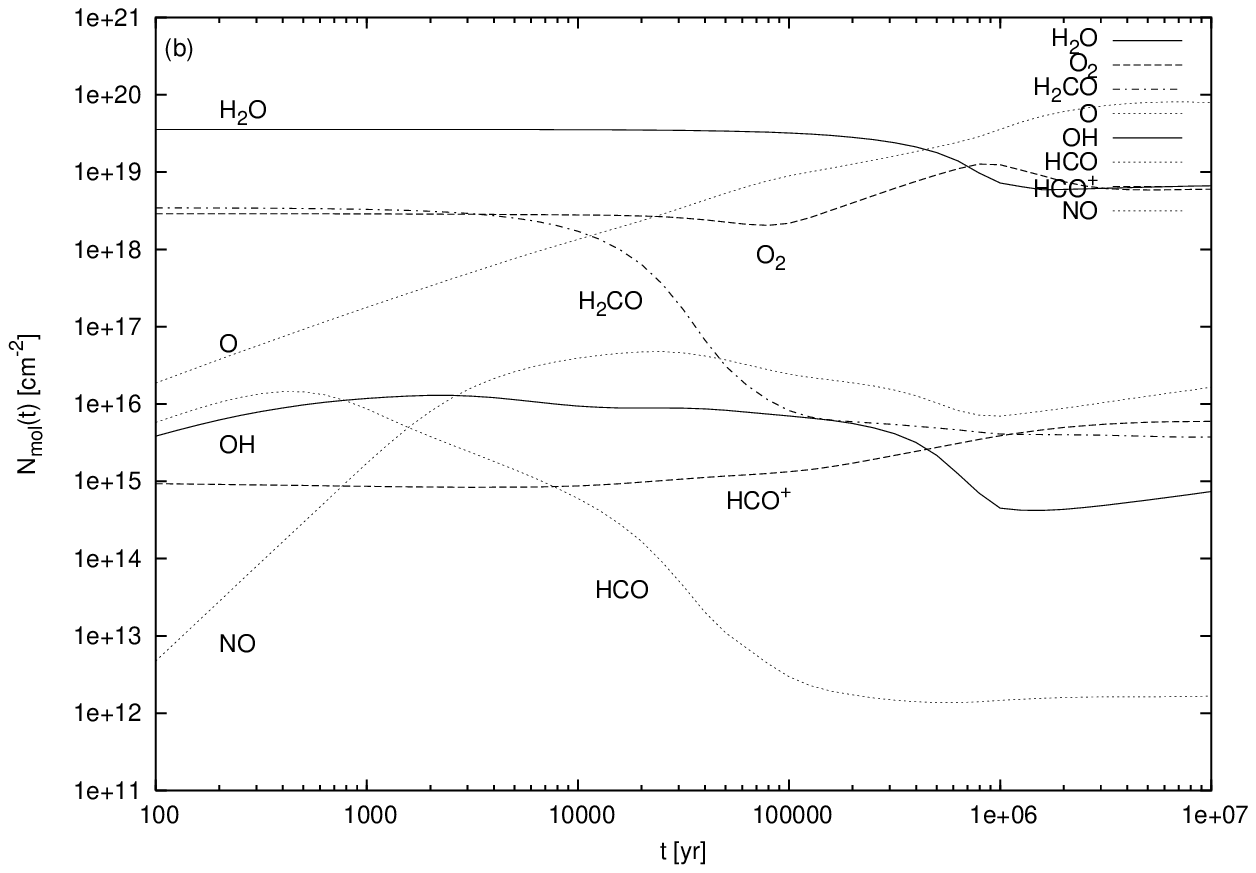}}
\resizebox{\hsize}{!}{\includegraphics{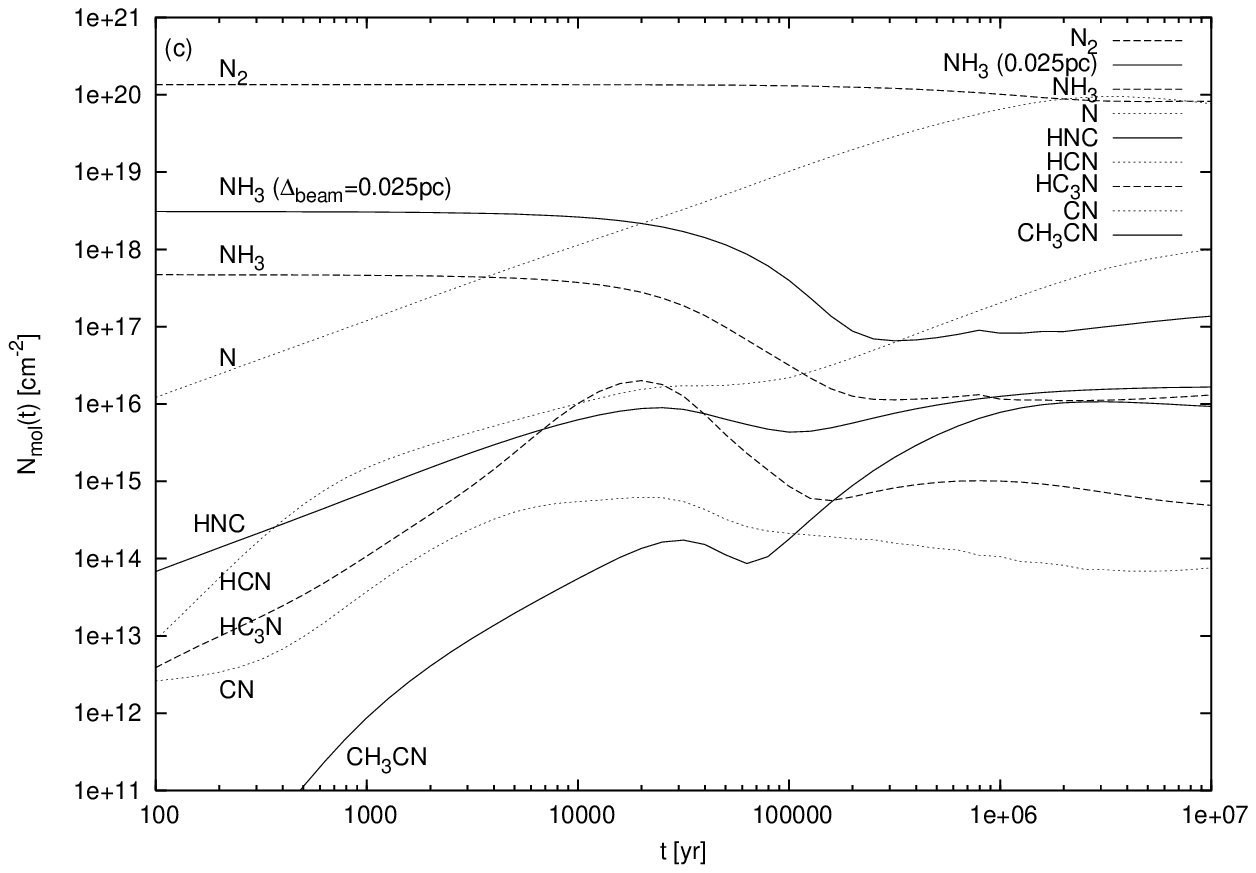}}
\end{figure}
\begin{figure}
\resizebox{\hsize}{!}{\includegraphics{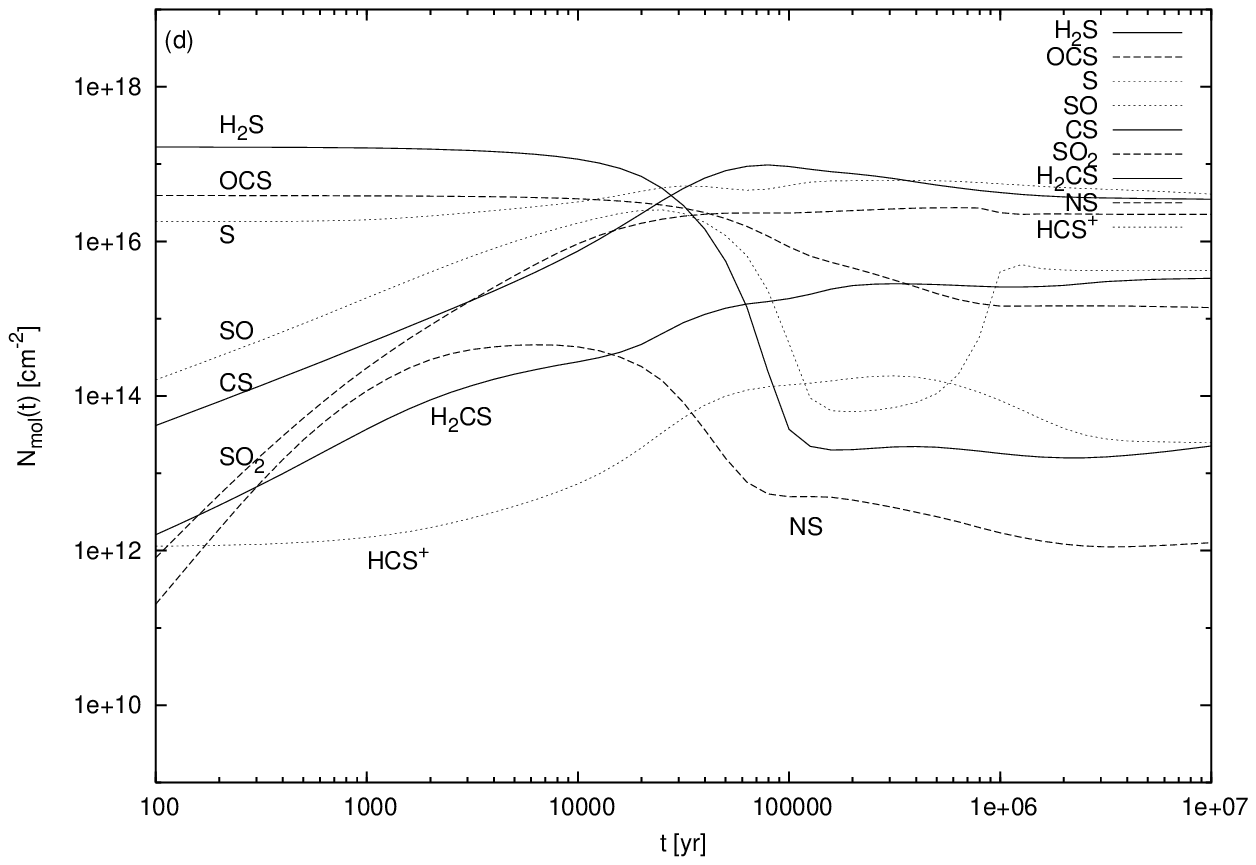}}
\resizebox{\hsize}{!}{\includegraphics{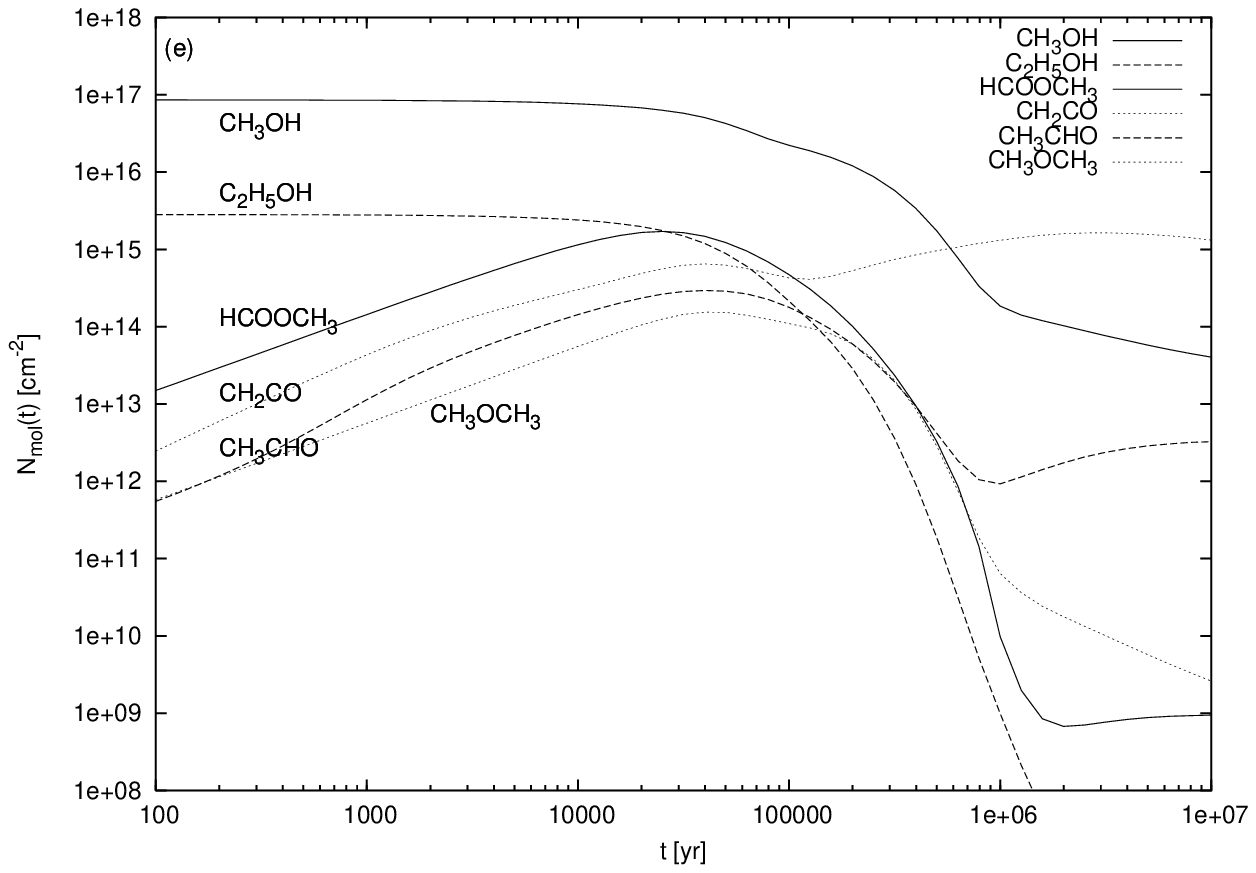}}
\caption{The time evolution of column densities of some
 molecules in the model of G34.3+0.15. Parent molecules are eventually
 destroyed to produce daughter species around the timescale of 10$^4$ yr.}
\end{figure}

The data at wavelengths from near- to mid-infrared can not be reproduced
by this model probably because of the inappropriate
assumption of the spherically symmetric density profile. For example, the 
optical depth of non-spherical cavities due to bipolar outflows would be
less than that of the dense envelopes so that if we were to observe the
inner region of the clumps through optically thin cavities,
namely, there would be more flux in the infrared 
(cf. Hatchell et al. 2000; van der Tak
et al. 2000a). Actually, the model with the column density of $N_{\rm
H}=10^{24}$ cm$^{-2}$ (1/10 of the above envelope model) and the core
radius of $r_{\rm c}=0.05$ pc, which is plotted with a dotted line 
in Fig. 1c, better represents the observational data in
the mid-infrared. Here,
the bolometric luminosity ratio contributed from the envelope
region with $N_{\rm H}=10^{25}$ cm$^{-2}$ to the cavity region with
$N_{\rm H}=10^{24}$ cm$^{-2}$ is assumed to be $20:1$ in order to fit
the observational data. This will mean that the
solid angle of the cavity will be much less than $4\pi$ and/or the
radiation flux in the clump around the central star is not
spherically symmetric in fact. Its density
and temperature profiles are also shown with dotted lines in Figs. 1a
and 1b. In the following subsections, we only use the model with the
column density of $N=10^{25}$ cm$^{-2}$ because most of the mass of the
clump will be contained in this dense envelope region rather than the
less dense, cavity region.

\subsection{Results of chemical calculation}

\subsubsection{Time evolution of molecular column densities}

\begin{figure}
\centering
\resizebox{\hsize}{!}{\includegraphics{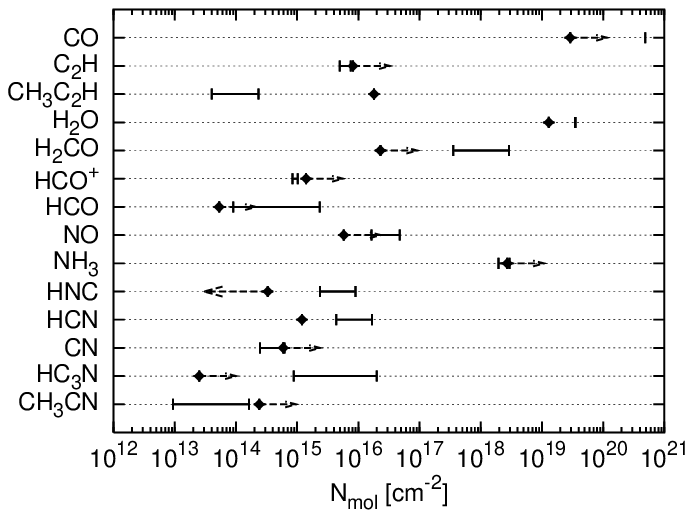}}
\resizebox{\hsize}{!}{\includegraphics{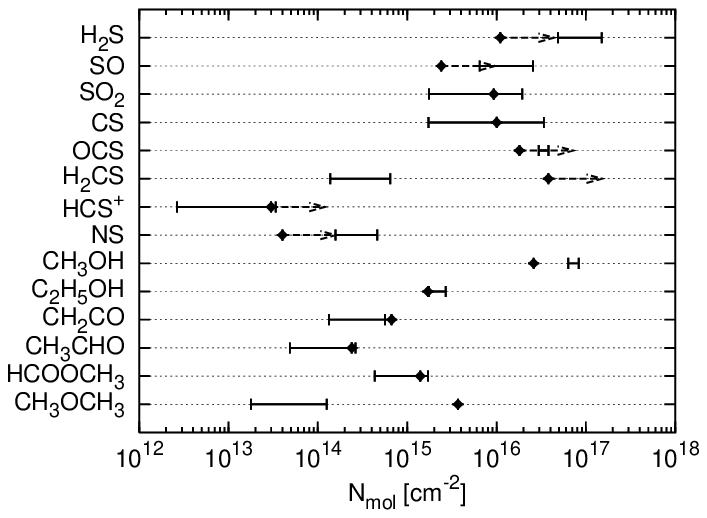}}
\caption{Comparison between the calculated molecular column densities
 from $3\times 10^3$ yr to $3\times 10^4$ yr (solid lines) and the
 observations towards G34.3+0.15 (diamonds plus dotted arrows if they
 are the lower or upper limits). The results are in good agreement with
 most of the observations.}
\end{figure}

Making use of the density and temperature structures obtained in the
previous subsection and the chemical model in Sect. 3, we simulate the time
evolution of the hot core G34.3+0.15. The resulting time evolution of
molecular column densities are shown in Fig.2. In order to compare these
with the observations of beam-averaged column densities, each atomic or
molecular column density is calculated as
\begin{equation}
N_i(t)=\dfrac{1}{\pi\Theta_{\rm beam}^2}\int_{0}^{\infty}2\pi r\tilde{N_i}(r,t)\exp\biggl(-\dfrac{r^2}{\Theta_{\rm beam}^2}\biggr)dr, \label{eq7}
\end{equation}
where $\tilde{N_i}(r,t)=\int_{-\infty}^{\infty}n(s)x_i(s,t)ds$ is the
column density of species $i$ along a line of sight at a distance $r$
from the clump centre. Here, $n(s)$ is the number density of hydrogen
nuclei, $x_i(s,t)$ the fractional abundance of species $i$ at time $t$, 
and $s$ the
depth along the line of sight. The scale size $\Theta_{\rm beam}$ is
simply taken as 0.25 pc which corresponds to the angular size of 13'' at
the distance of 3.7kpc, based on the half power beam width of 
the James Clerk Maxwell Telescope (JCMT) at 345 GHz (cf. Thompson et
al. 1999). While the data of most molecular lines we use in this paper
are from single radio telescope observations such as the JCMT, the
NH$_3$ data is from the Very Large Array observation whose synthesized
beam size is about 1''3 (Heaton et al. 1989). Thus, we calculate the
NH$_3$ column density with $\Theta_{\rm beam}=0.025$ pc as well as
with $\Theta_{\rm beam}=0.25$ pc for comparison.

\begin{table}
\caption[]{Molecular column densities: centre vs. off-centre.}\label{T2}
$$ \begin{array}{p{0.17\linewidth}|cr|cr} \hline 
 & \multicolumn{4}{|c}{{\rm Molecular\ column\ density}\ [{\rm cm}^{-2}]} \\ \cline{2-5}
 & \multicolumn{2}{|c|}{\rm Centre} & \multicolumn{2}{|c}{\mbox{Off-centre}} \\ \cline{2-5}
 Species & {\rm Model^{\mathrm{a}}} & {\rm Observation} & {\rm Model^{\mathrm{a}}} & {\rm Observation} \\ \hline
H$_2$ & 3.6\times 10^{24} & & 2.0\times 10^{24} & \\
CO & 4.9\times 10^{20} & >2.9\times 10^{19} & 2.6\times 10^{20} & >1.3\times 10^{19} \\
C$_2$H & 7.4\times 10^{15} & >8.2\times 10^{15} & 4.9\times 10^{15} & >2.9\times 10^{14} \\
CH$_3$C$_2$H & 1.1\times 10^{14} & 1.8\times 10^{16} & 1.9\times 10^{13} & \\
H$_2$O & 3.5\times 10^{19} & 1.3\times 10^{19} & 4.5\times 10^{18} & \\
H$_2$CO & 1.7\times 10^{18} & >2.3\times 10^{16} & 3.0\times 10^{17} & >1.3\times 10^{14} \\
HCO$^+$ & 8.7\times 10^{14} & >1.4\times 10^{15} & 6.0\times 10^{14} & >2.3\times 10^{13} \\
HCO & 6.4\times 10^{14} & >5.3\times 10^{13} & 1.3\times 10^{14} & \\
NO & 3.9\times 10^{16} & >5.8\times 10^{15} & 1.2\times 10^{16} & \\
NH$_3$ & 3.7\times 10^{17} & & 3.9\times 10^{16} & \\
NH$_3$ $^{\mathrm{b}}$ & 2.6\times 10^{18} & >2.7\times 10^{18} & & \\
HNC & 6.2\times 10^{15} & <3.3\times 10^{14} & 8.6\times 10^{14} & \\
HCN & 1.0\times 10^{16} & 1.2\times 10^{15} & 1.3\times 10^{15} & >2.3\times 10^{13} \\
CN & 5.5\times 10^{14} & >5.9\times 10^{14} & 3.9\times 10^{14} & >9.5\times 10^{12} \\
HC$_3$N & 1.0\times 10^{16} & >2.5\times 10^{13} & 9.7\times 10^{14} & \\
CH$_3$CN & 5.5\times 10^{13} & >2.4\times 10^{14} & 6.5\times 10^{12} & \\
H$_2$S & 1.2\times 10^{17} & >1.1\times 10^{16} & 1.9\times 10^{16} & \\
SO & 1.7\times 10^{16} & >2.4\times 10^{15} & 2.3\times 10^{15} & \\
SO$_2$ & 9.3\times 10^{15} & 9.3\times 10^{15} & 9.9\times 10^{14} & >2.0\times 10^{14} \\
CS & 7.5\times 10^{15} & 1.0\times 10^{16} & 1.4\times 10^{15} & >5.3\times 10^{13} \\
OCS & 3.6\times 10^{16} & >1.8\times 10^{16} & 3.4\times 10^{15} & \\
H$_2$CS & 2.8\times 10^{14} & >3.8\times 10^{16} & 7.9\times 10^{13} & \\
HCS$^+$ & 7.3\times 10^{12} & >3.0\times 10^{13} & 3.6\times 10^{12} & \\
NS & 4.3\times 10^{14} & >4.0\times 10^{13} & 8.7\times 10^{13} & \\
CH$_3$OH & 7.7\times 10^{16} & 2.6\times 10^{16} & 6.9\times 10^{15} & >3.6\times 10^{14} \\
C$_2$H$_5$OH & 2.4\times 10^{15} & 1.7\times 10^{15} & 2.3\times 10^{14} & \\
CH$_2$CO & 3.0\times 10^{14} & 6.7\times 10^{14} & 1.3\times 10^{14} & \\
CH$_3$CHO & 1.4\times 10^{14} & 2.4\times 10^{14} & 1.5\times 10^{13} & \\
HCOOCH$_3$ & 1.1\times 10^{15} & 1.4\times 10^{15} & 9.2\times 10^{13} & \\
CH$_3$OCH$_3$ & 5.6\times 10^{13} & 3.7\times 10^{15} & 5.0\times 10^{12} & \\ \hline
\end{array}
$$ 
\begin{list}{}{}
\item[$^{\mathrm{a}}$] Calculated results at $10^4$ yr.
\item[$^{\mathrm{b}}$] In the case of $\Theta_{\rm beam}=0.025$ pc.
\end{list}
\end{table}

From Fig.2 we can see that parent molecules, such as H$_2$O, NH$_3$,
H$_2$S, and CH$_3$OH, which are expected to be formed on grain surfaces,
are relatively stable for timescales of at least 10$^4$ yr in this
model. However, they are eventually
destroyed by ionized species or atomic hydrogen to produce daughter
species (see Sect. 4.2.3 for the details)
as has been suggested in the previous studies (e.g., review by Millar
1997). Thus, the observed large abundances of both parent and daughter
molecules will mean that the ages of the chemically rich hot cores are
about 10$^4$ yr. In particular, we compare the results of our calculation with
observations in Fig.3, which shows that the calculated molecular column
densities from $3\times 10^3$ yr to $3\times 10^4$ yr (solid lines) are in
good agreement with most of those observed toward the hot core
G34.3+0.15 (diamonds plus dotted arrows if they are the lower or upper
limits), consistent with the result of Millar et al. (1997). 
Lower limits in this Table have been derived from optically thick 
transitions.  The calculated molecular column densities, with the exception
of  CH$_3$CCH, HNC,
H$_2$CS, and CH$_3$OCH$_3$, satisfy the observational lower limits or fit
the observations to within a factor 4 at most. 
Discrepancies may be resolved in a number of ways. Dimethyl ether, 
CH$_3$OCH$_3$, may in fact be 
a mantle molecule, while the gas-phase chemistry of H$_2$CS is not at all 
well understood.  Unlike H$_2$CO, for which efficient gas-phase synthetic
pathways have been studied in the laboratory, the analogous reactions which
might form thioformaldehyde do not occur.   The chemistry of HNC at low
temperatures is well understood; the discrepancy between model and observation 
may imply that additional high-temperature loss mechanisms which are not
included in the model are at work. Finally, the methyl acetylene, CH$_3$CCH,
column density is about a factor of 100 less than that observed by Macdonald 
et al. (1996) from their 330--360 GHz spectral line scan. However, it is much 
closer to the value of 8.2 $\times$ 10$^{14}$ cm$^{-2}$ determined from the 
2 and 3 mm
line surveys of Kim et al. (2000), similar to the value derived by Hatchell 
et al.(1998a) 20$\arcsec$ offset from the hot core.  Finally, we note that
the formation of C$_3$H$_5^+$, the precursor ion to CH$_3$CCH, involves
ions and neutrals related to CH$_4$, C$_2$H$_2$ and C$_2$H$_4$. An increase
in the abundances of each of these parents by a factor of a few would also
help resolve the discrepancy noted above. 
The observational data are taken from Millar et al. (1997) (see also
Macdonald et al. 1996 and Hatchell et al. 1998a) and Ikeda et al. (2001).

\subsubsection{Radial profiles of molecular abundances}

\begin{figure}
\centering
\resizebox{\hsize}{!}{\includegraphics{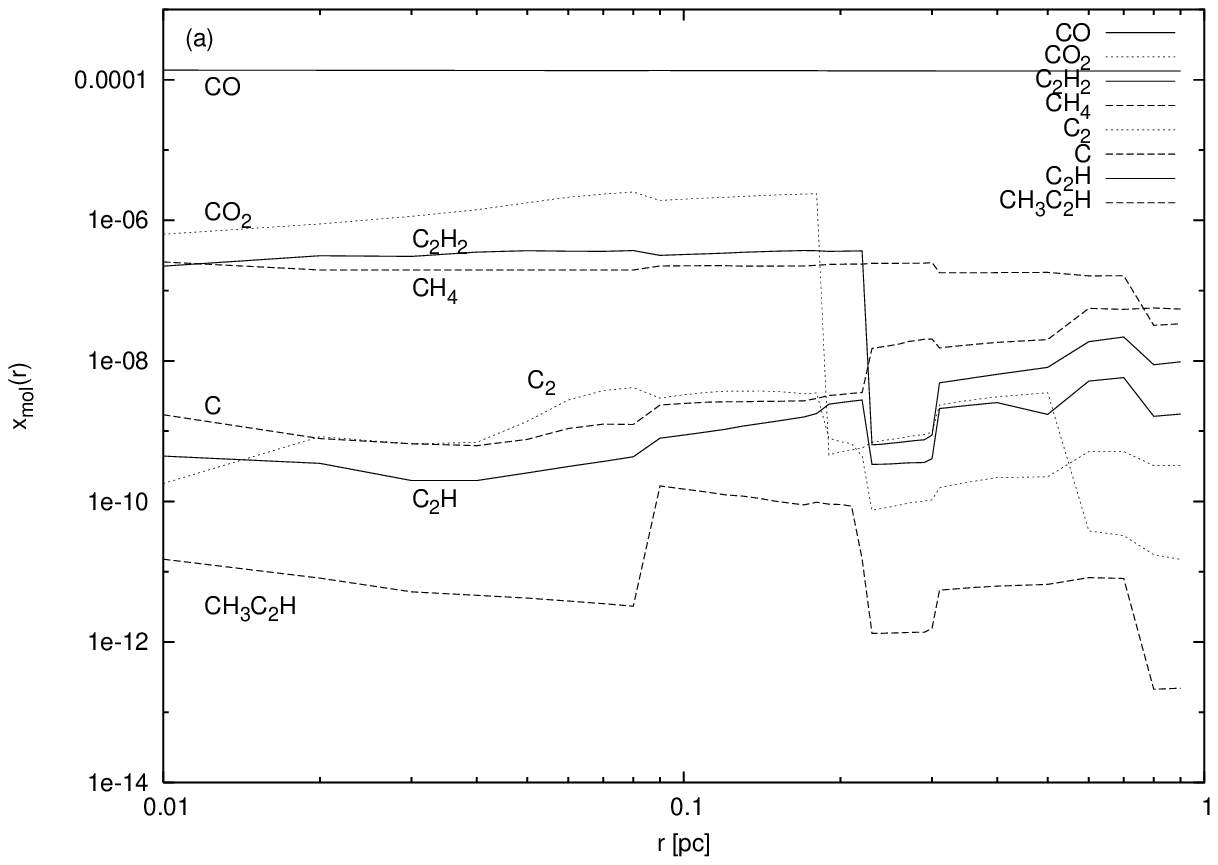}}
\resizebox{\hsize}{!}{\includegraphics{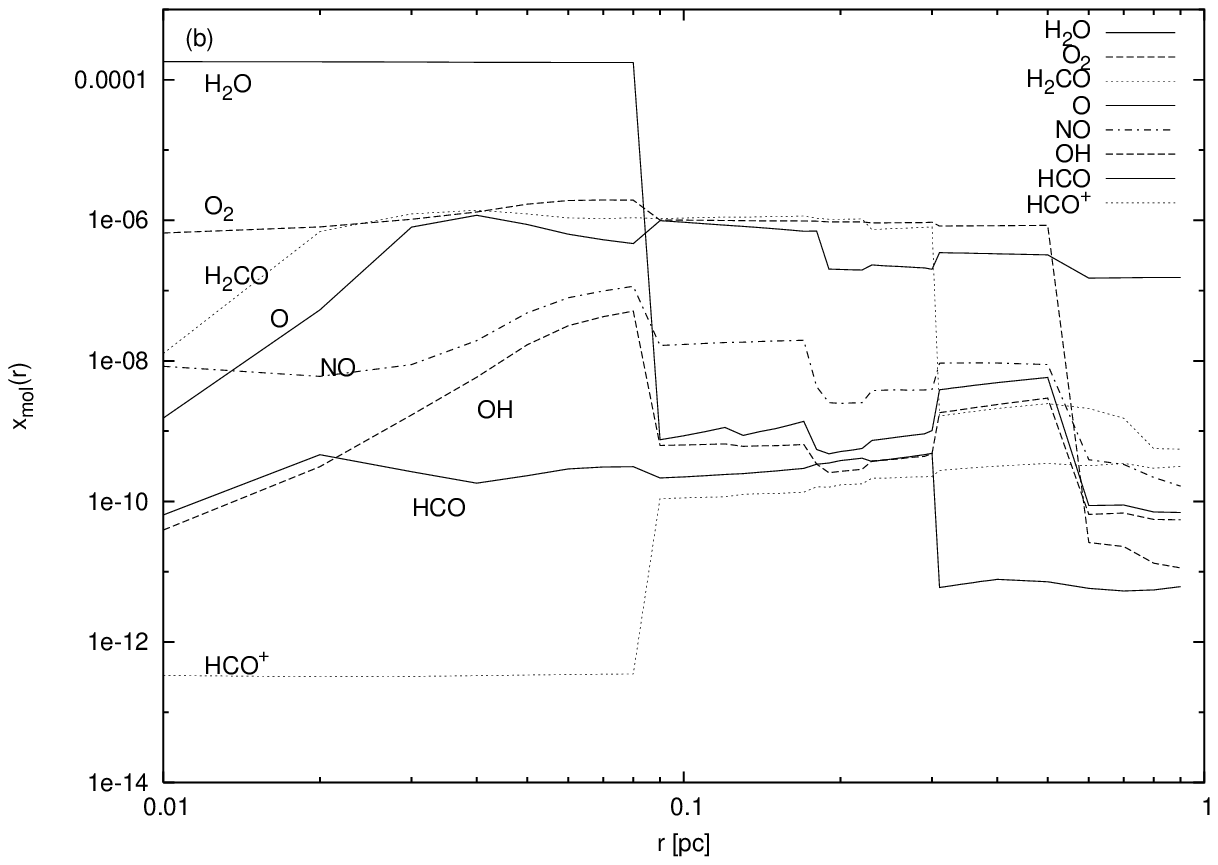}}
\resizebox{\hsize}{!}{\includegraphics{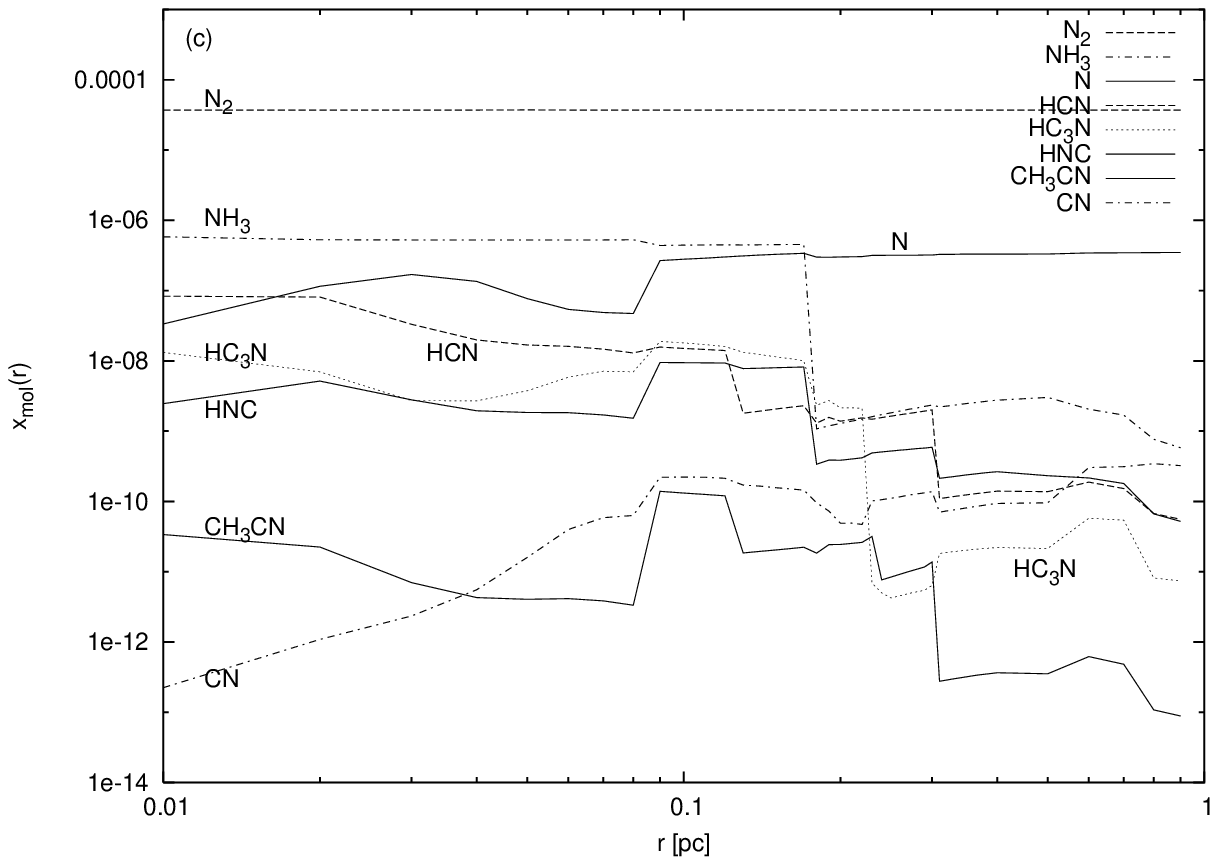}}
\end{figure}
\begin{figure}
\resizebox{\hsize}{!}{\includegraphics{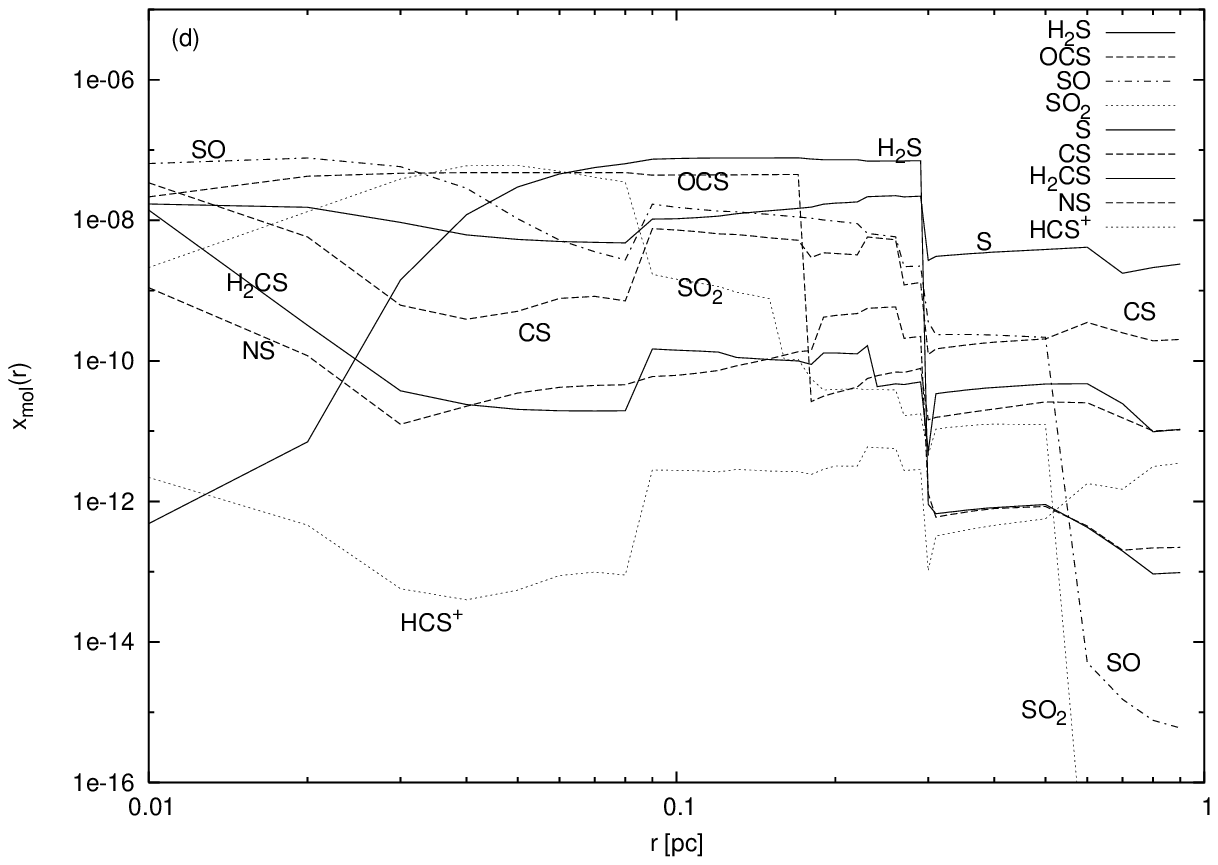}}
\resizebox{\hsize}{!}{\includegraphics{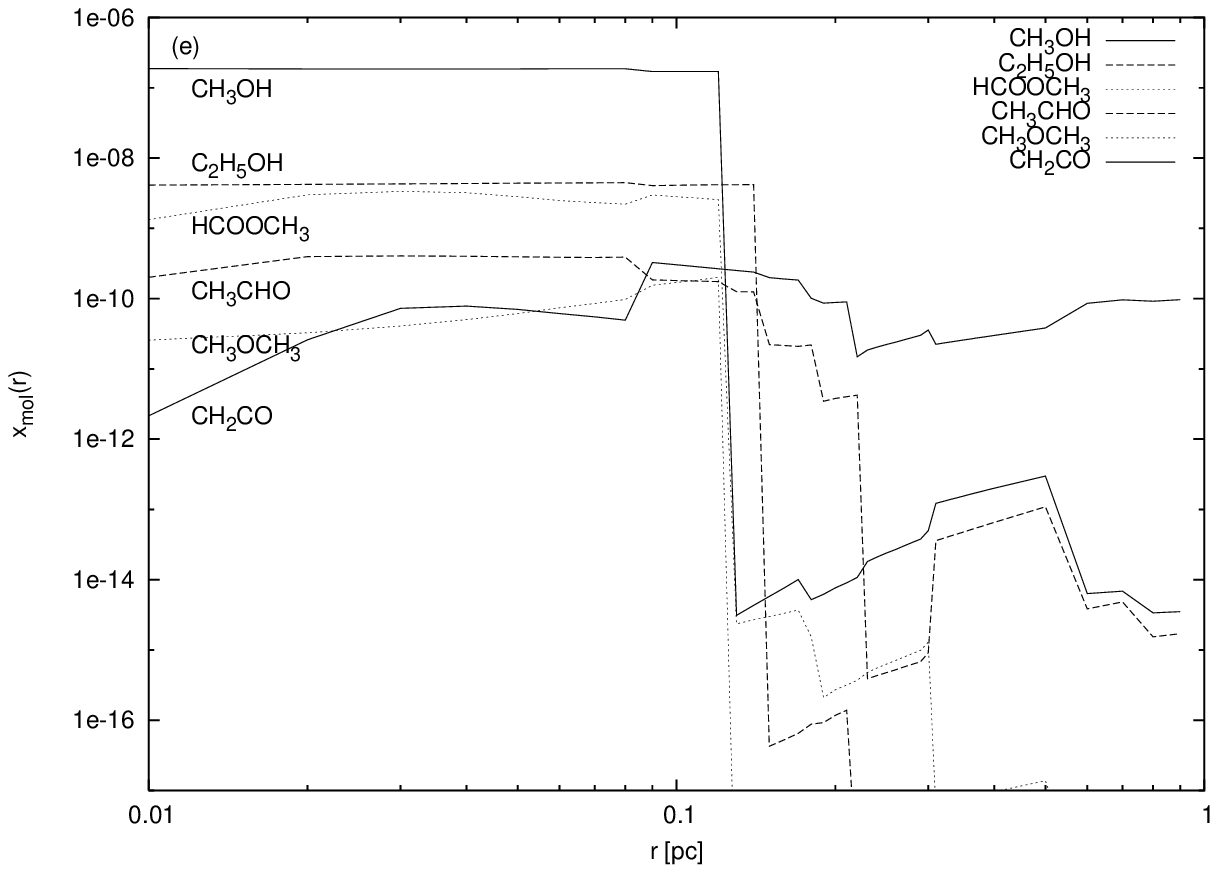}}
\caption{The radial profiles of some molecular abundances of the model
 of G34.3+0.15 at $10^4$ yr. Some parent, daughter and other related
 species have dramatic changes at radii related to the grain mantle
 evaporation.}
\end{figure}

In order to see the effects of the temperature dependent injection of
molecules from dust grains on hot core chemistry, we present
the radial profiles of molecular abundances at $10^4$ yr in Fig.4. We
can see from the figures that the abundances of many parent
molecules injected from grain mantles, such as C$_2$H$_2$, CO$_2$,
H$_2$O, H$_2$S, and CH$_3$OH, change dramatically at $r_{{\rm inj}, i}$,
inside which the molecule $i$ can be easily evaporated. Inside 
$r_{{\rm inj}, i}$, most of the parents are difficult to destroy. 
Nevertheless, given their large initial abundances, even a slow
destruction rate is large enough to produce appreciable abundances
of daughter products.
Thus, the radial abundance profiles of the parent molecules influence
those of daughter and other related species (see Sect. 4.2.3 for
details). Abundance changes of some molecules along the radial
direction in the clumps around young massive stars have been suggested
by observational studies. For instance, the observations of the CH$_3$CN
rotational transition lines indicate that they are emitted from the
inner compact hot region of the clumps, while the observed CH$_3$OH
lines suggest that the gas in some clumps consists of at least two
components, one of which is hot and abundant in CH$_3$OH, and the other
is cooler and less abundant (Hatchell et al. 1998a; Millar \& Hatchell
1998; van der Tak et al. 2000b). In addition, the observational
abundances of the vibration-rotation lines of C$_2$H$_2$, HCN, H$_2$O,
and CO$_2$ are found to increase with the excitation temperature
(Lahuis \& van Dishoeck 2000; Boonmann et al. 2000).
Detailed radiative transfer calculations will be needed to compare the
results of our theoretical model with these observations. We note that
abundance jumps similar to those predicted have been observed in the
low-mass `hot core' around IRAS 16293-2422 (Sch\"oier et al. 2002).

Our results show that a different chemistry operates in the
surrounding cooler envelopes from that in the central hot regions where
icy mantles can be evaporated, as has been indicated (e.g., Millar et
al. 1997; Doty et al. 2002). In order to see the quantitative
differences comparable with observations, we calculate the molecular
column densities off-centred from the clump centre by a distance $r_{\rm
off}$, making use of Eq. (\ref{eq7}) with the distance from the 
clump centre, $r'=r-r_{\rm off}$ replacing $r$.
The results at $t=10^4$ yr are listed in Table \ref{T2}. Parameters of
$\Theta_{\rm beam}=0.25$ pc and $r_{\rm off}=0.4$ pc are adopted for
comparing the results with observations toward the halo associated
with G34.3+0.15 by Thompson et al. (1999), which are also included in Table
\ref{T2}. For comparison, the molecular column densities calculated
around the clump centre by Eq. (\ref{eq7}) with $t=10^4$ yr are
also listed together with the observations which are the same as those
in Fig.3. The results of the model calculations are consistent with most
of the observations. The ratio of the calculated molecular column
densities around the clump centre to those off-centred range roughly from
a factor 2, when the radial abundances of the species are almost
constant, to one order of magnitude, in the case of parent or daughter species
which are hardly generated by the gas-phase reactions without grain
mantle evaporation.

\subsubsection{Detailed chemistry}

In this subsection, we discuss some of the hot core chemistry,
mainly focusing on the effects of the injection of parent molecules
from grain mantles on the radial abundance profiles of daughter and
other related species.

Fig.2(a) and Fig.2(b) show the time evolution of column densities of the
carbon- and oxygen-bearing species, while Fig.4(a) and Fig.4(b)
represent the radial profiles of their abundances at $t=10^4$ yr.
The results related to oxygen chemistry can be interpreted based on
the detailed analysis by Charnley (1997).
Water is the main parent molecule among oxygen-bearing species and
mainly destroyed by HCO$^+$ to produce H$_3$O$^+$ in this model.
This process keeps the HCO$^+$ abundance low where H$_2$O is injected
from the grain mantles and is consistent
with the result of Doty et al (2002). Hydroxyl molecules are formed by
the dissociative recombination of H$_3$O$^+$, so that the OH abundance 
changes discontinuously at the H$_2$O injection radius, $r_{{\rm inj,
H}_2{\rm O}}$. One of the
destruction processes of carbon monoxide is the reaction with He$^+$,
which makes atomic oxygen as well as atomic carbon.
On the other hand, O and OH react with molecular hydrogen to reproduce
water as O $\stck{{\rm H}_2}$ OH $\stck{{\rm H}_2}$ H$_2$O at high
temperature. Thus, the abundances of O and OH are suppressed in
the inner hot region. 
These behaviours are also
consistent with the results of Doty et al (2002). The reaction between
OH and N generates NO, which makes the radial abundance profile of NO 
similar to that of OH. Molecular oxygen is not only injected from the
grain mantles but also synthesized from O and OH. HCO is principally
produced by the destruction of H$_2$CO by OH, so that the HCO abundance
jump appears at $r_{{\rm inj, H}_2{\rm CO}}$.
C$_3$H$_4$ is chiefly formed initiated from C$_2$H$_2$ as C$_2$H$_2$
$\stck{{\rm HCO}^+}$ C$_2$H$_3^+$ $\stck{{\rm CH}_4}$ C$_3$H$_5^+$
$\stck{e}$ C$_3$H$_4$ in this model. Therefore, the C$_3$H$_4$ abundance
increases over a radial distance between $r_{{\rm inj, H}_2{\rm O}}$, 
inside which the HCO$^+$ abundance is suppressed, and  
$r_{{\rm inj, C}_2{\rm H}_2}$.

Fig.2(c) and Fig.4(c) are the time evolution of column densities and the
radial abundance profiles of nitrogen-bearing species, whose detailed
hot core chemistry have
analysed by Millar et al. (1997) and Rodgers \& Charnley (2001). 
The nitrogen molecule is the most dominant nitrogen-bearing species in the
initial gas in this model, and is converted into N and N$^+$, through reaction 
with He$^+$. Ammonia is one of the possible nitrogen-bearing mantle species
and is destroyed by ionized molecules to produce NH$_2$. Meanwhile,
it is formed from gas-phase reactions initiated from N$^+$ as
N$^+$ $\stck{{\rm H}_2}$ NH$^+$ $\stck{{\rm H}_2}$ NH$_2^+$ $\stck{{\rm
H}_2}$ NH$_3^+$ $\stck{{\rm H}_2}$ NH$_4^+$ $\stck{e}$ NH$_3$.
The radial abundance profile of HCN has
two apparent discontinuous jumps in this model. One is at the radius
where $\tau_{\rm acc, HCN}=\tau_{\rm evap, HCN}$ and the other is at
$r_{{\rm inj, H}_2{\rm CO}}$ as H$_2$CO is one of the main precursors of
CH$_2$, which reacts with N to produce HCN. 
The major production reaction of HNC is that between NH$_2$ and C. Thus,
the HNC abundance jumps at $r_{{\rm inj, NH}_3}$.
CH$_3$CN is mainly produced by the radiative association between HCN and
CH$_3^+$, so that the radial abundance profile of CH$_3$CN resembles
that of HCN. The CH$_3$CN abundance is suppressed inside $r_{{\rm
inj, H}_2{\rm O}}$ because H$_3^+$, which helps produce CH$_3^+$ from 
its reaction with 
CH$_3$OH, decreases in abundance as it preferentially transfers its proton to 
the abundant water molecule.
HC$_3$N is generated from N as N $\stck{{\rm C}_3{\rm H}}$ C$_3$N
$\stck{{\rm XH}^+}$ HC$_3$N$^+$ $\stck{{\rm H}_2}$ H$_2$C$_3$N$^+$
$\stck{\rm e}$ HC$_3$N. C$_3$H is a product of C$_2$H$_2$, which causes
the HC$_3$N abundance jump at $r_{{\rm inj, C}_2{\rm H}_2}$. CN is also
chiefly formed from H$_2$C$_3$N$^+$ inside $r_{{\rm inj, C}_2{\rm H}_2}$,
but the jump does not appear in the CN abundance because CN can be
synthesized by neutral-neutral reactions such as C+NO outside $r_{{\rm
inj, C}_2{\rm H}_2}$. The CN abundance is suppressed in the inner hot
region due to the loss by a reaction with H$_2$ which
creates HCN.

Fig.2(d) and Fig.4(d) represent the time evolution of column densities
and the radial abundance profiles of sulfur-bearing species,
respectively. Their basic behaviour can be explained following the
analyses by Millar et al. (1997) and Charnley (1997). Hydrogen sulfide
is a possible sulfur-bearing grain mantle species, and
can be easily destroyed by atomic hydrogen especially in the inner hot
region because of the small activation energy barrier of this reaction. 
This drives 
the sulfur chemistry in hot cores as H$_2$S $\stck{\rm H}$ SH
$\stck{\rm H}$ S $\stck{{\rm O}_2}$ SO $\stck{\rm OH}$ SO$_2$. 
Thus, the SO abundance is lowered inside $r_{{\rm inj, H}_2{\rm O}}$,
where the OH abundance is high, while the SO$_2$ abundance drops outside
$r_{{\rm inj, H}_2{\rm O}}$. Also, both the SO and SO$_2$ abundances
change dramatically at
$r_{{\rm inj, O}_2}$. NS is produced by the reaction between SH and N, so
that the NS abundance changes discontinuously at $r_{{\rm inj, H}_2{\rm
S}}$. CS is mainly generated as SO $\stck{\rm C}$ CS in the inner
region, while HCS$^+$ $\stck{\rm e}$ CS in the outer region. On the
other hand, HCS$^+$ is synthesized by the proton transfer process to CS
in the inner region, whereas by the reaction between CS$^+$ and H$_2$ 
in the outer region. Therefore, the radial abundance profiles of HCS$^+$
and CS resemble each other.
OCS is injected from grain mantles in this model in order to represent
the observed high abundance of OCS molecular lines, and based on the
observations of ice absorption features by Palumbo et al. (1997) (see
also discussions in Hatchell et al. 1998b). 

\begin{figure}
\centering
\resizebox{\hsize}{!}{\includegraphics{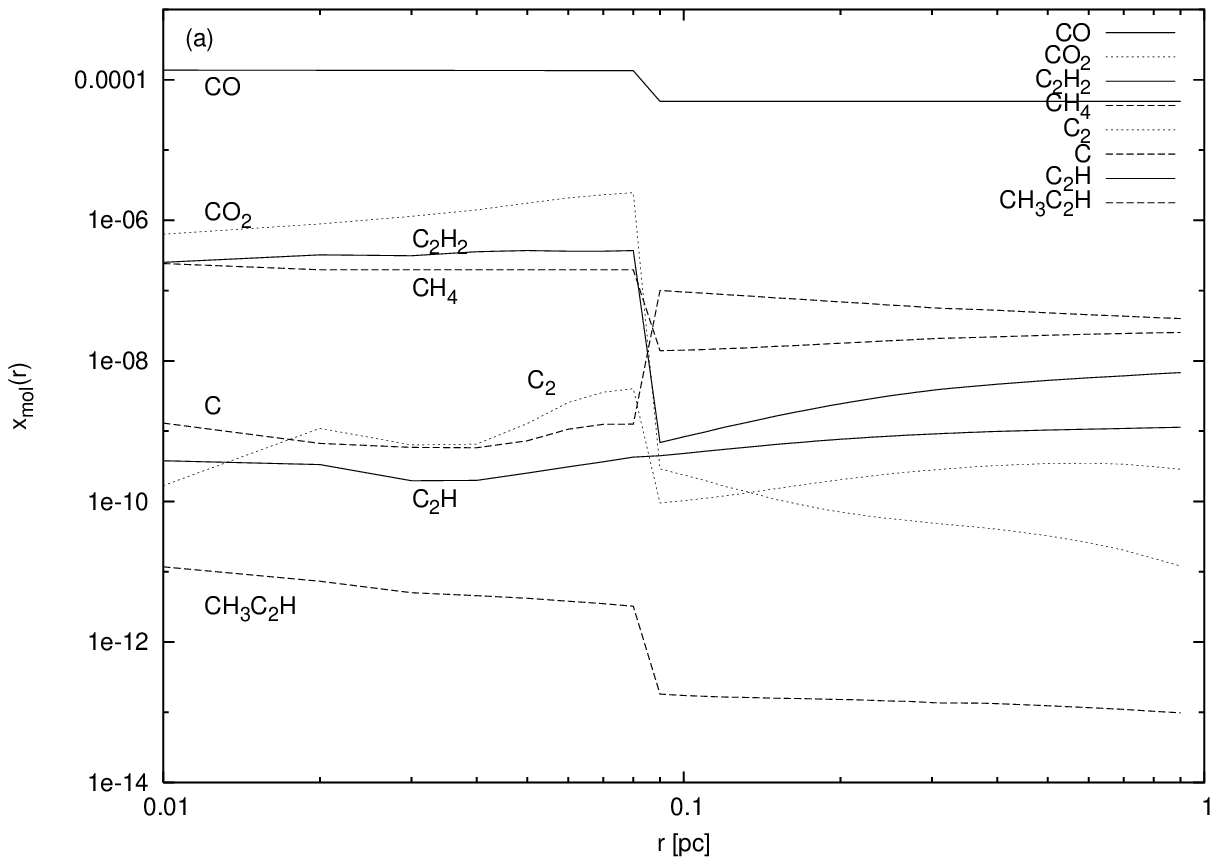}}
\resizebox{\hsize}{!}{\includegraphics{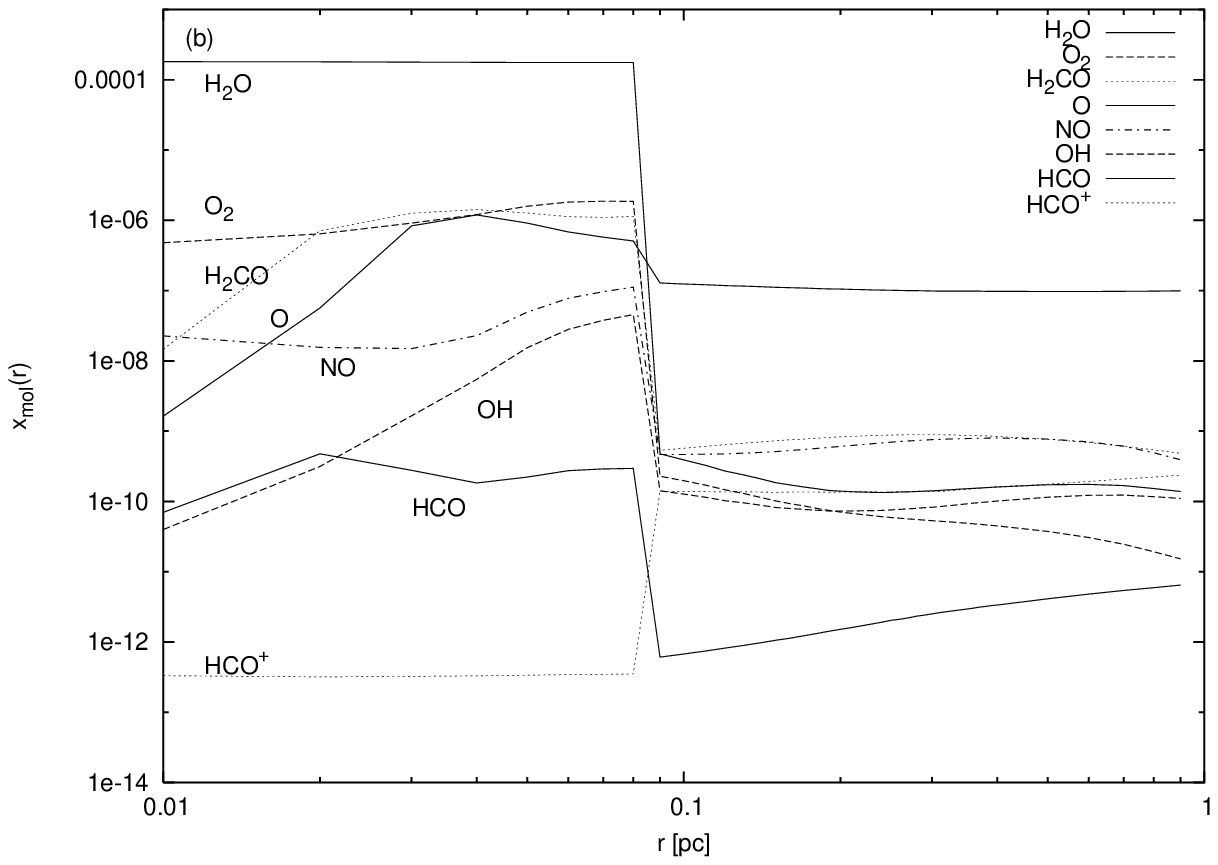}}
\resizebox{\hsize}{!}{\includegraphics{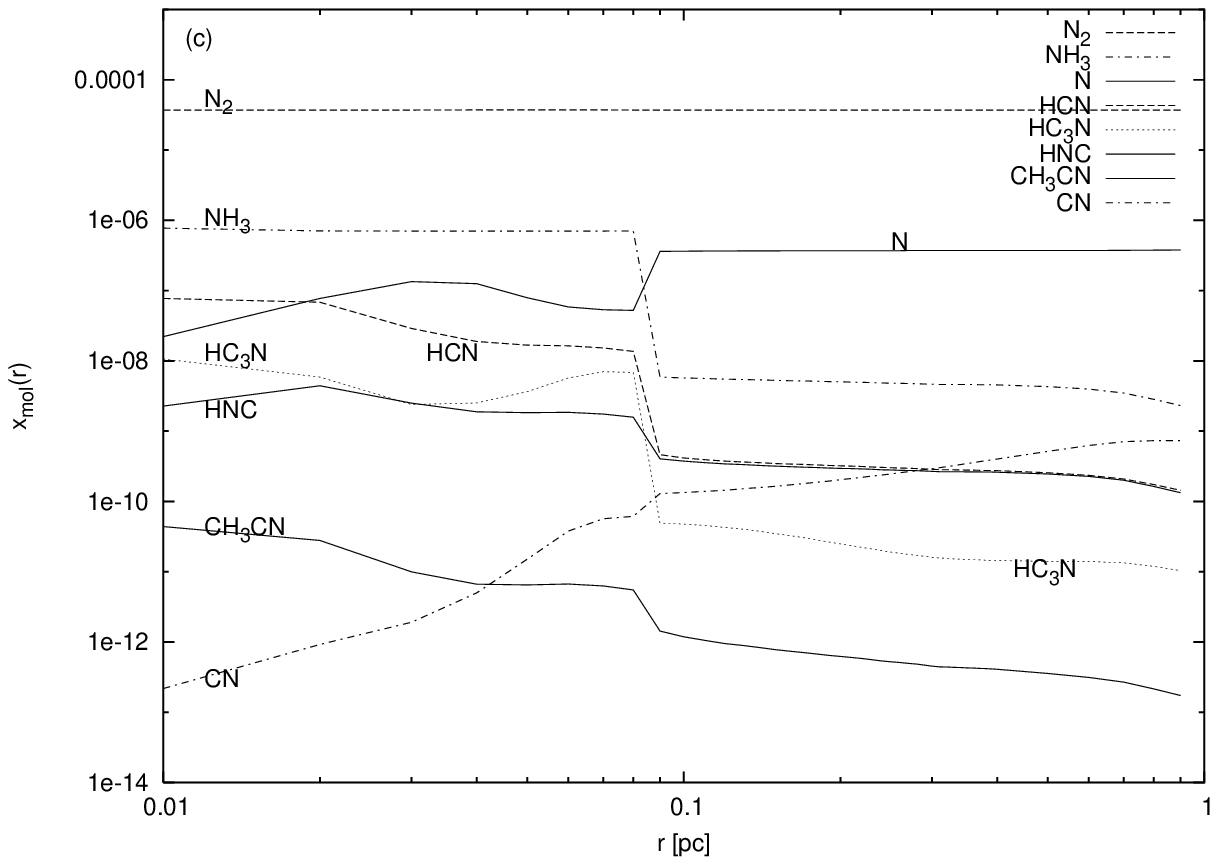}}
\end{figure}
\begin{figure}
\resizebox{\hsize}{!}{\includegraphics{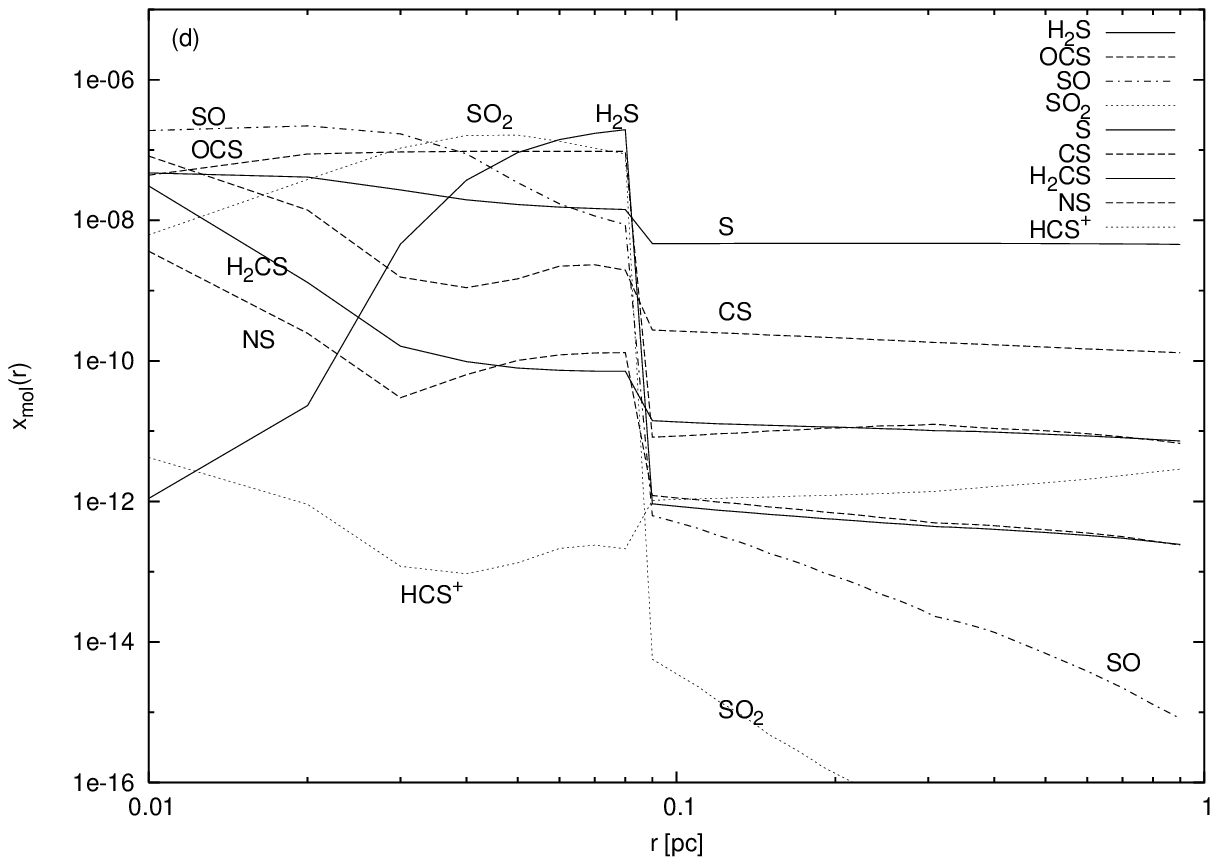}}
\resizebox{\hsize}{!}{\includegraphics{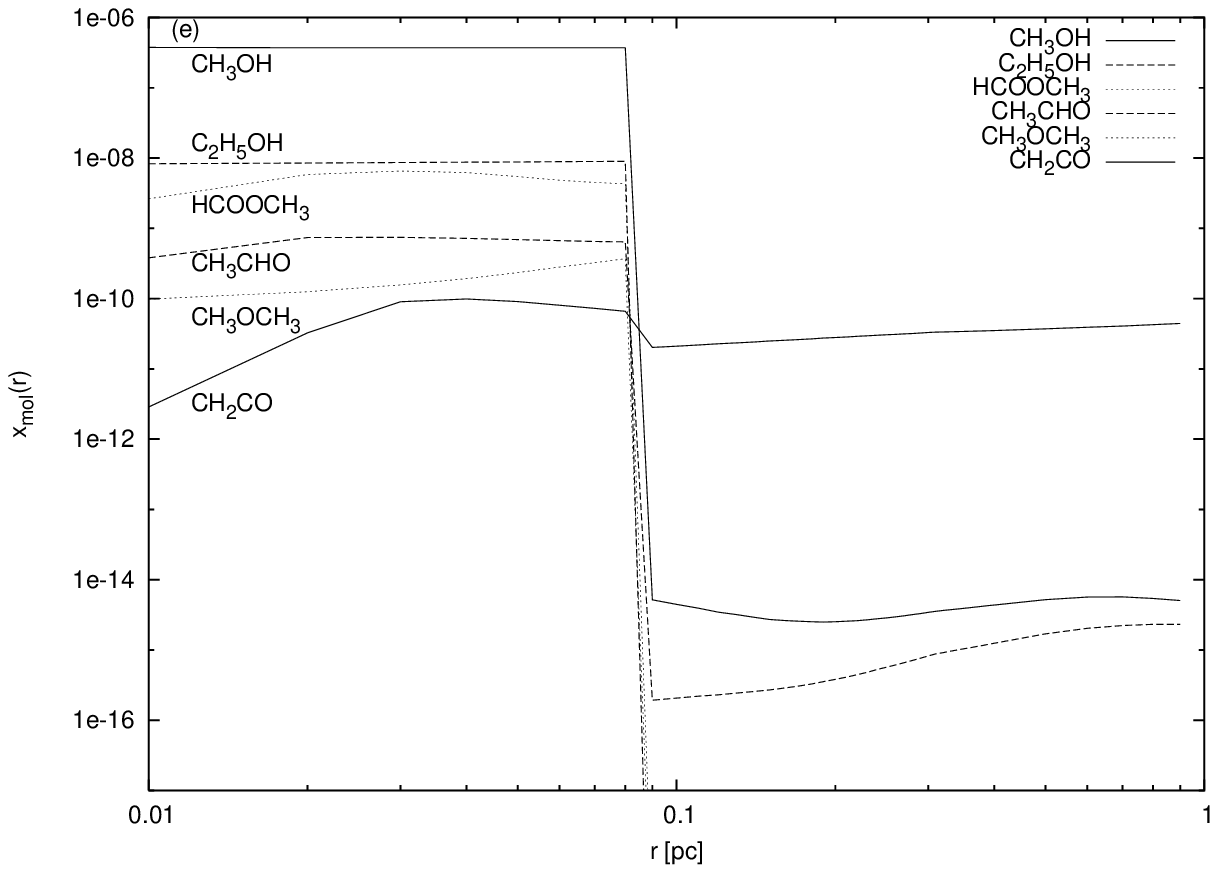}}
\caption{The same as Fig.4 but with the trapping of mantle molecules in
 water ice. Large differences appear in the radial distributions
of some molecules compared to those in Fig.4.}
\end{figure}
%

Fig.2(e) and Fig.4(e) show the time evolution of column densities
and the radial abundance profiles of large oxygen-bearing species, which
can be interpreted according to the analyses by Millar et al. (1991) and
Charnley et al. (1992). Methanol is one of the possible large
oxygen-bearing grain mantle molecules
and is protonated by H$_3^+$. The product of the protonation,
CH$_3$OH$_2^+$, leads to the formation of daughter molecules as
CH$_3$OH$_2^+$ $\stck{{\rm H}_2{\rm CO}}$ HCOOCH$_4^+$ $\stck{\rm e}$
HCOOCH$_3$ and CH$_3$OH$_2^+$ $\stck{{\rm CH}_3{\rm OH}}$
C$_2$H$_6$OH$^+$ $\stck{\rm e}$ (CH$_3$)$_2$O. Thus, abundance jumps of
HCOOCH$_3$ and (CH$_3$)$_2$O appear at $r_{{\rm inj, CH}_3{\rm OH}}$. 
CH$_3$CHO is mainly made by the reaction between atomic oxygen and
C$_2$H$_5$ which is generated from C$_2$H$_6$, so that the CH$_3$CHO
abundance changes dramatically at $r_{{\rm inj, C}_2{\rm H}_6}$. CH$_2$CO
is chiefly formed from the reactions of C$_2$H$_3$ $\stck{\rm
O}$ CH$_2$CO and CH$_3^+$ $\stck{\rm CO}$ CH$_3$CO$^+$ $\stck{\rm e}$
CH$_2$CO inside $r_{{\rm inj, C}_2{\rm H}_4}$, while from the latter
reaction otherwise. 
In this model C$_2$H$_5$OH is assumed to be created on dust grains in
order to reproduce the observations of C$_2$H$_5$OH molecular lines
(cf. Millar et al. 1997, Rodgers \& Charnley 2001, Ikeda et al. 2001).

\section{Discussion}

\subsection{Trapping of mantle molecules in water ice}

\begin{table}
\caption[]{Initial abundances of mantle molecules for Model B}\label{T3}
$$ \begin{array}{p{0.11\linewidth}c|p{0.12\linewidth}c|p{0.12\linewidth}c} \hline 
 Species & {\rm Abundance} & Species & {\rm Abundance} & Species & {\rm Abundance}  \\ \hline
 C$_2$H$_2$ & 5.0\ 10^{-7} & CO$_2$ & 3.0\ 10^{-6} & O$_2$ & 1.0\ 10^{-6} \\
 CH$_4$ & 2.0\ 10^{-7} & H$_2$CO & 2.0\ 10^{-6} & H$_2$O & 2.8\ 10^{-4} \\
 C$_2$H$_4$ & 1.0\ 10^{-8} & CH$_3$OH & 4.0\ 10^{-7} & NH$_3$ & 8.0\ 10^{-7} \\
 C$_2$H$_6$ & 1.0\ 10^{-8} & C$_2$H$_5$OH & 1.0\ 10^{-8} & H$_2$S & 3.0\ 10^{-7} \\
 & & & & OCS & 1.0\ 10^{-7} \\ \hline
\end{array}
$$ 
\end{table}
%
\begin{table}
\caption[]{Molecular column densities of Model B and observations.}\label{T4}
$$ \begin{array}{p{0.13\linewidth}|cr|p{0.17\linewidth}|cr} \hline 
 & \multicolumn{2}{|c|}{N_{\rm mol}\ [{\rm cm}^{-2}]} & & \multicolumn{2}{|c}{N_{\rm mol}\ [{\rm cm}^{-2}]} \\ \cline{2-3} \cline{5-6}
 Species & {\rm Model\ B^{\mathrm{a}}} & \multicolumn{1}{c|}{\rm Obs.} & Species & {\rm Model\ B^{\mathrm{a}}} & \multicolumn{1}{c}{\rm Obs.} \\ \hline
CO & 2.0\ 10^{20} & >2.9\ 10^{19} & H$_2$S & 2.4\ 10^{16} & >1.1\ 10^{16} \\
C$_2$H & 3.2\ 10^{15} & >8.2\ 10^{15} & SO & 9.7\ 10^{15} & >2.4\ 10^{15} \\
CH$_3$C$_2$H & 1.3\ 10^{12} & 1.8\ 10^{16} & SO$_2$ & 2.3\ 10^{16} & 9.3\ 10^{15} \\
H$_2$O & 3.5\ 10^{19} & 1.3\ 10^{19} & CS & 1.5\ 10^{15} & 1.0\ 10^{16} \\
H$_2$CO & 2.3\ 10^{17} & >2.3\ 10^{16} & OCS & 1.9\ 10^{16} & >1.8\ 10^{16} \\ 
HCO$^+$ & 5.5\ 10^{14} & >1.4\ 10^{15} & H$_2$CS & 1.8\ 10^{14} & >3.8\ 10^{16} \\ 
HCO & 6.4\ 10^{13} & >5.3\ 10^{13} & HCS$^+$ & 5.6\ 10^{12} & >3.0\ 10^{13} \\
NO & 1.6\ 10^{16} & >5.8\ 10^{15} & NS & 5.0\ 10^{13} & >4.0\ 10^{13} \\
NH$_3$ $^{\mathrm{b}}$ & 2.6\ 10^{18} & >2.7\ 10^{18} & CH$_3$OH & 7.4\ 10^{16} & 2.6\ 10^{16} \\ 
HNC & 1.3\ 10^{15} & <3.3\ 10^{14} & C$_2$H$_5$OH & 1.8\ 10^{15} & 1.7\ 10^{15} \\ 
HCN & 5.0\ 10^{15} & 1.2\ 10^{15} & CH$_2$CO & 1.3\ 10^{14} & 6.7\ 10^{14} \\
CN & 1.3\ 10^{15} & >5.9\ 10^{14} & CH$_3$CHO & 1.3\ 10^{14} & 2.4\ 10^{14} \\ 
HC$_3$N & 1.1\ 10^{15} & >2.5\ 10^{13} & HCOOCH$_3$ & 1.0\ 10^{15} & 1.4\ 10^{15} \\ 
CH$_3$CN & 3.4\ 10^{12} & >2.4\ 10^{14} & CH$_3$OCH$_3$ & 5.5\ 10^{13} & 3.7\ 10^{15} \\ \hline 
\end{array}
$$ 
\begin{list}{}{}
\item[$^{\mathrm{a}}$] Calculated results at $10^4$ yr.
\item[$^{\mathrm{b}}$] In the case of $\Theta_{\rm beam}=0.025$ pc.
\end{list}
\end{table}

It is known that water is one of the most abundant ice component in
dense clouds. Some observations of ice absorption features in infrared
spectra towards young stellar objects have suggested that a certain
amount of CO is frozen 
in H$_2$O-rich matrices (e.g., review by Allamandola et al. 1999), and
laboratory experiments have shown that solid CO is able to diffuse into
the porous structure of H$_2$O and be trapped in it (Collings et al. 2003).
These studies suggest that mantle molecules other than CO also could be
trapped in water ice. Thus, in this subsection we examine the effects of
this trapping on the chemical structure of the hot core G34.3+0.15. In order to
see the extreme case, we put the injection radius of each mantle
molecule as $r_{{\rm inj},i}\equiv r_{{\rm inj, H}_2{\rm O}}$, and
assume that it is completely trapped in H$_2$O matrices outside $r_{{\rm
inj, H}_2{\rm O}}$, except CO, for which we assume that $x({\rm CO})=0.3x({\rm
H}_2{\rm O})$ is trapped outside $r_{{\rm inj, H}_2{\rm
O}}$ based on observations (e.g., Chiar et al. 1998; cf. Doty et
al. 2002). As a result of
the time-dependent chemical calculation described in Sect. 3, we get the 
molecular column densities of the hot core consistent with 
most of the observations at around $10^4$ yr (Table 4).
The initial condition listed in Table \ref{T3} is used in this calculation,
which is slightly different from that in Table 1, in order to fit the
results to the observations. Although there are few differences in the
integrated molecular column densities, we find that very different
structure appear in the radial profiles of molecular abundances at around
$10^4$ yr between
the results in Sect. 4 (Fig.4: Model A) and those obtained from this
calculation (Fig.5: Model B). Especially the differences are remarkable
in the following molecules: (1) the binding energies of the molecules or
their parent species are low, that is, the injection radii listed in
Table \ref{T1} are large, such as H$_2$CO, HCO, H$_2$S, and SO, or (2)
daughter molecules whose production are suppressed where H$_2$O is
evaporated from grain mantle, such as CH$_3$C$_2$H and CH$_3$CN. For
comparison, the resulting radial
abundance profiles of SO and CH$_3$CN from Model A (dotted lines) and B
(solid lines) are plotted in the linear radial scales in Fig.6.
Detailed observations of radial profiles of molecular
abundances may make it possible to examine what fraction of molecules is
trapped in water ice in cooler region of hot cores.

\begin{figure}
\centering
\resizebox{\hsize}{!}{\includegraphics{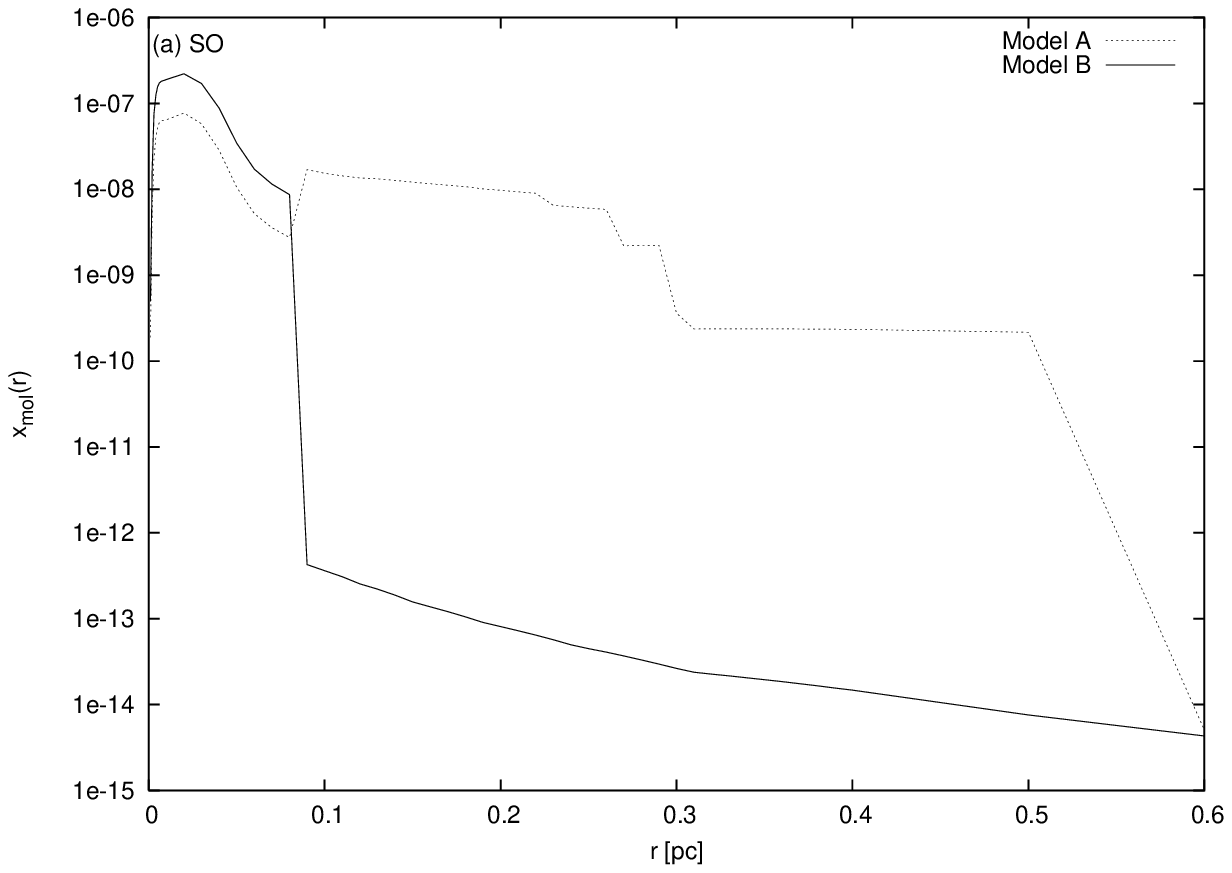}}
\resizebox{\hsize}{!}{\includegraphics{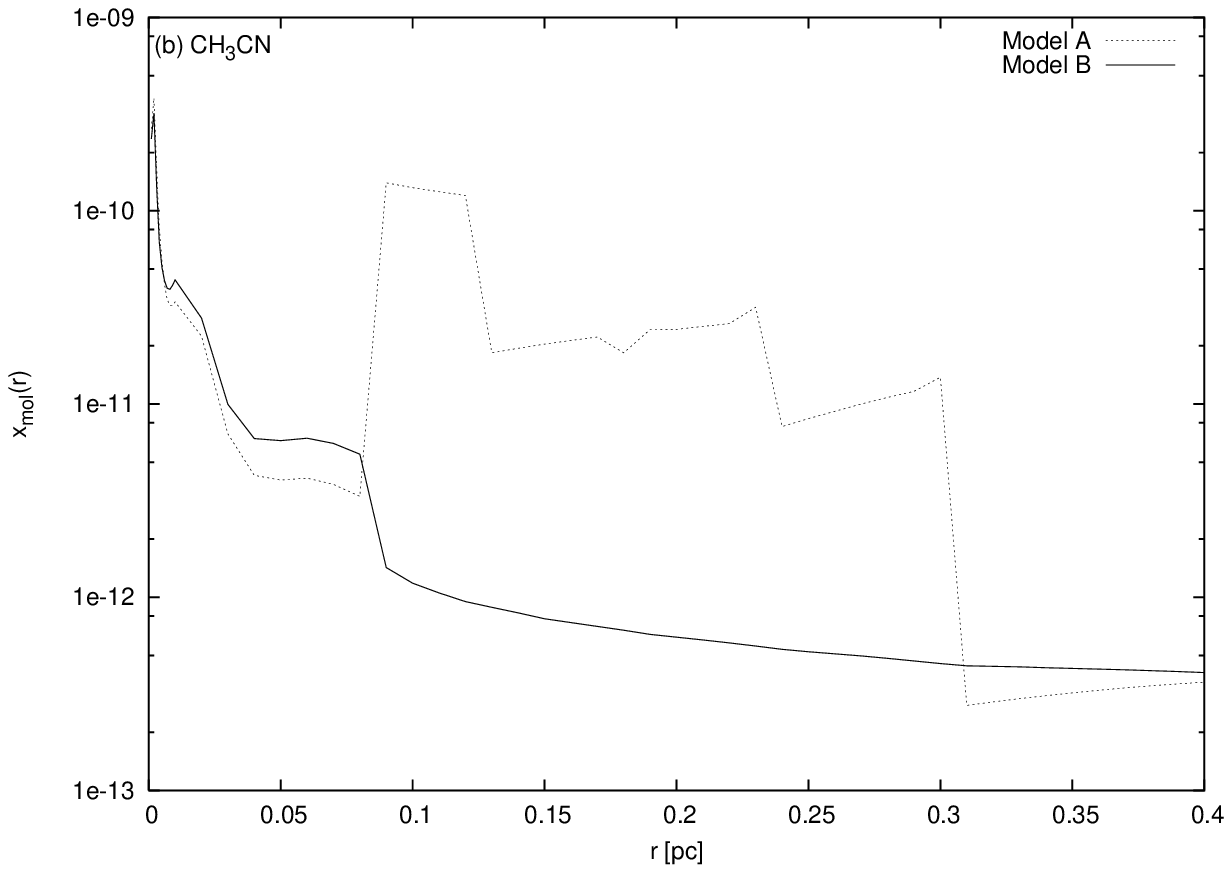}}
\caption{The comparison of the radial abundance profiles of (a) SO and
 (b) CH$_3$CN in the linear radial scales between the models 
 without (dotted lines: Model A) and with (solid
 lines: Model B) the trapping of mantle
 molecules in water ice. We can find the difference in the 
 location of jump in (a) and that the abundance is low through the clump
 in Model B in (b).}
\end{figure}

\subsection{Dependence of chemical structure on
density profile -- General model}

The comparison between the results of chemical calculation and the
observations of various kinds of molecular lines will be useful to
constrain the physical condition of massive star forming regions (e.g.,
Viti \& Williams 1999). The density profile of hot molecular cores is
one of the possible tracers of the evolution process
just before and after massive star formation.
In this section we will investigate the dependences of the chemical
structure of hot cores on the density profile to discuss the physical
condition of massive-star-forming clumps.

\begin{figure}
\centering
\resizebox{\hsize}{!}{\includegraphics{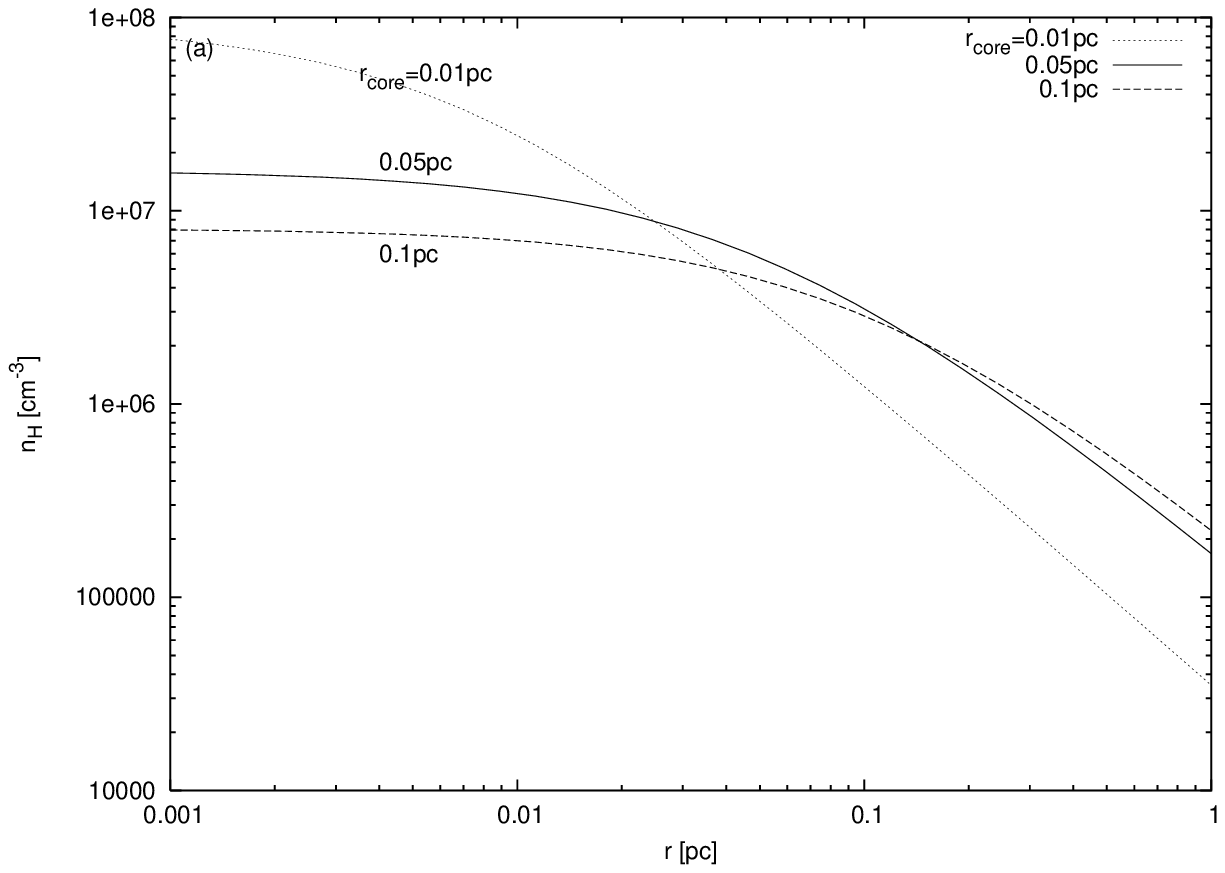}}
\resizebox{\hsize}{!}{\includegraphics{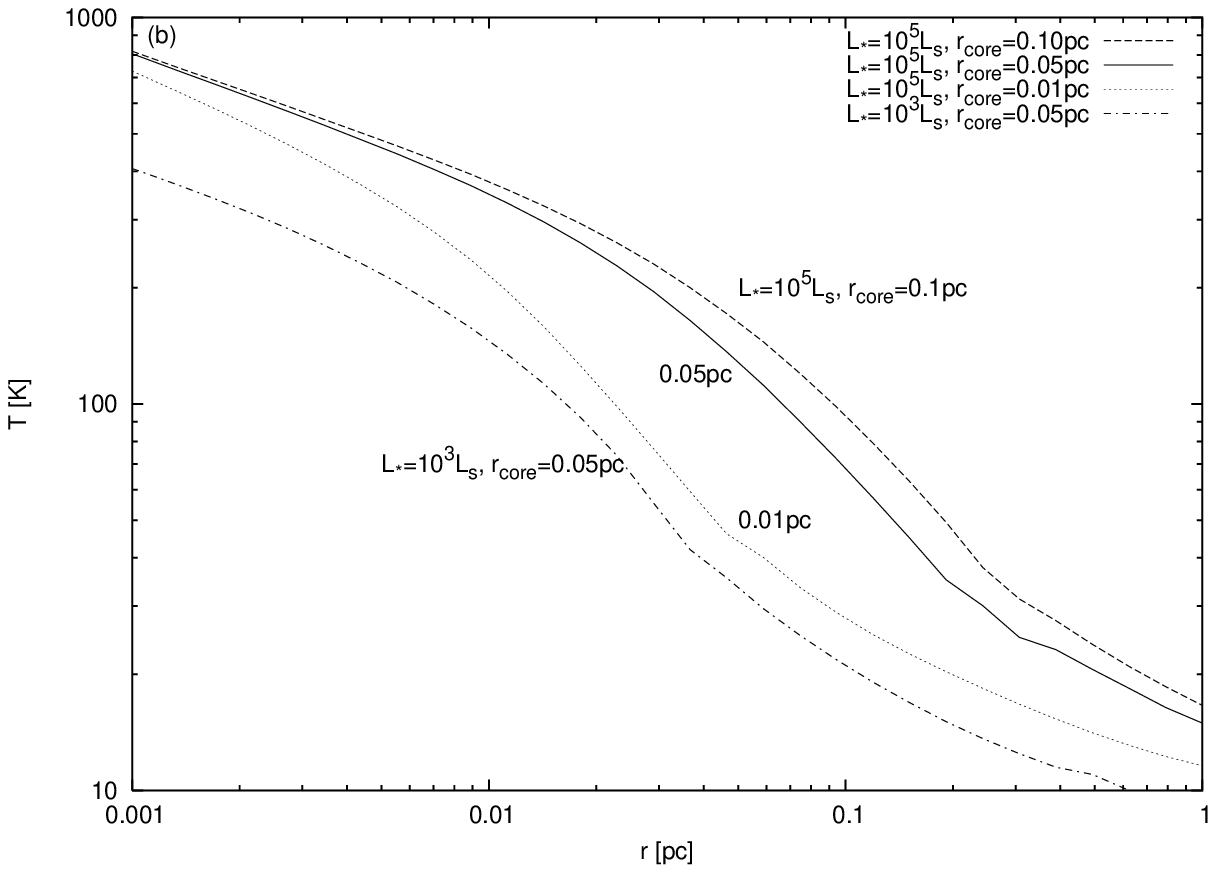}}
\end{figure}
\begin{figure}
\resizebox{\hsize}{!}{\includegraphics{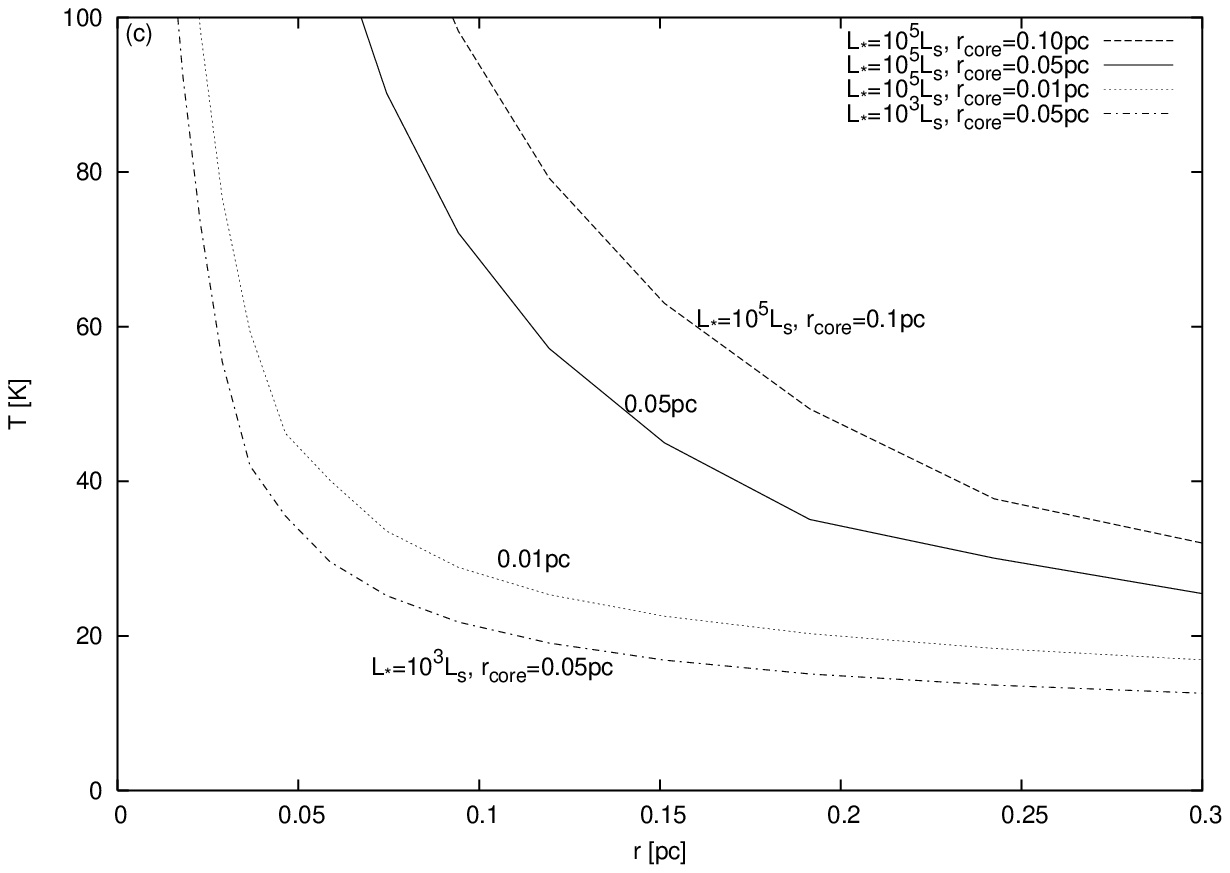}}
\caption{(a) Density and (b) temperature profiles in the logarithmic
 scales for the models with $L_*=10^5L_{\odot}$ and $r_{\rm c}=$
 0.01 (dotted lines), 0.05 (solid lines), and 0.1 pc (dashed lines). The
 temperature profiles directly reflect the difference in the density
 profiles. (c) Temperature profiles in the linear scales represent
 clearly the dependence on the density models around the temperatures of
 grain mantle evaporation. Profiles of the model with
 $L_*=10^3L_{\odot}$ and $r_{\rm c}=$ 0.05 pc (dot-dashed lines) are
 also plotted for comparison.}
\end{figure}

The chemical structures of three different types of molecular clupms are
investigated here. Their density profiles are set by Eq. (\ref{eq1})
with core radii of $r_{\rm c}=$ 0.01, 0.05, and 0.1 pc, and
plotted in Fig.7(a) with dotted, solid, and dashed lines, respectively. 
Making use of these density profiles, we calculate the temperature
profiles by solving the radiative transfer equation as described in
Sect. 2.2. 
The resulting temperature distributions are shown in Fig.7(b). We can
see from the figure that the temperature profiles directly reflect the
difference in the density profiles. This is because the temperature profiles
of hot cores are determined by the balance between the absorption of
radiation from the central star and the reemission of the radiation by dust
grains in the core as described by Eq. (\ref{eq2}).  The
inner regions of the cores are optically thick enough to
the radiation from the central stars so that the photons from the stars
diffuse via interactions with the surrounding dusty material. If we
simply assume that the dust opacity depends on the frequency of radiation
as $\kappa_{\nu}\propto \nu^n$ and the density profile of a clump is set
as $\rho\propto r^{-p}$, for example, the temperature profile of the
clump illuminated by a central star is derived as $T\propto
r^{-(p+1)/(4-n)}$ in the optically thick limit from the equation of
local radiative equilibrium and the radiative transfer equation (e.g.,
Adams \& Shu 1985). Meanwhile, since the evaporation process of grain
mantle species is very sensitive to the temperature as shown in
Sect. 3, it is to be expected that the density profiles affect the radial
abundance profiles of molecules in hot cores via the 
temperature profiles. In order to see the dependence on the density
models clearly, the temperature profiles in the region of grain mantle
evaporation are plotted with linear scales in Fig.7(c).

The luminosity of the central star is another factor that influences
the temperature profile of the clump. As we can see from Fig.7(b) and
(c), the temperature distribution of the clump whose core radius is
$r_{\rm c}=0.05$ pc and whose central star luminosity is
$L_*=10^3L_{\odot}$ (dot-dashed lines) is similar to that of the model
with $r_{\rm c}=0.01$ pc and $L_*=10^5L_{\odot}$. The difference in
luminosity will be, however, distinguishable by observing the 
spectral energy distribution. The hydrogen
column density will not have very much effect on the temperature structure 
in the optically thick region because basically $T(r)$ depends not
on the absolute density but on the steepness of the density profile as we
have shown in the above example. The optically thick region spreads
(shrinks) if the hydrogen column density is high (low) so that it will
affect the temperature profile at the outer radii.

\begin{figure}
\centering
\resizebox{\hsize}{!}{\includegraphics{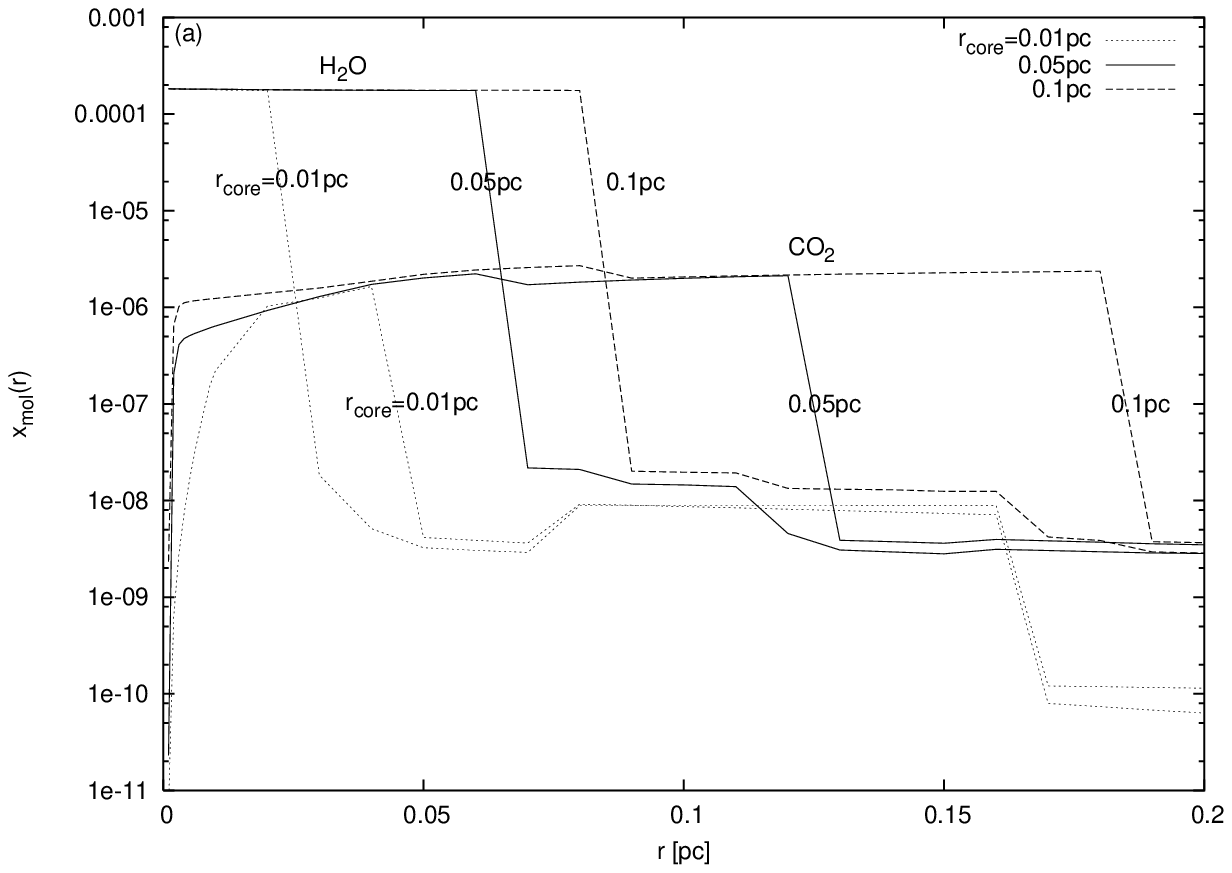}}
\resizebox{\hsize}{!}{\includegraphics{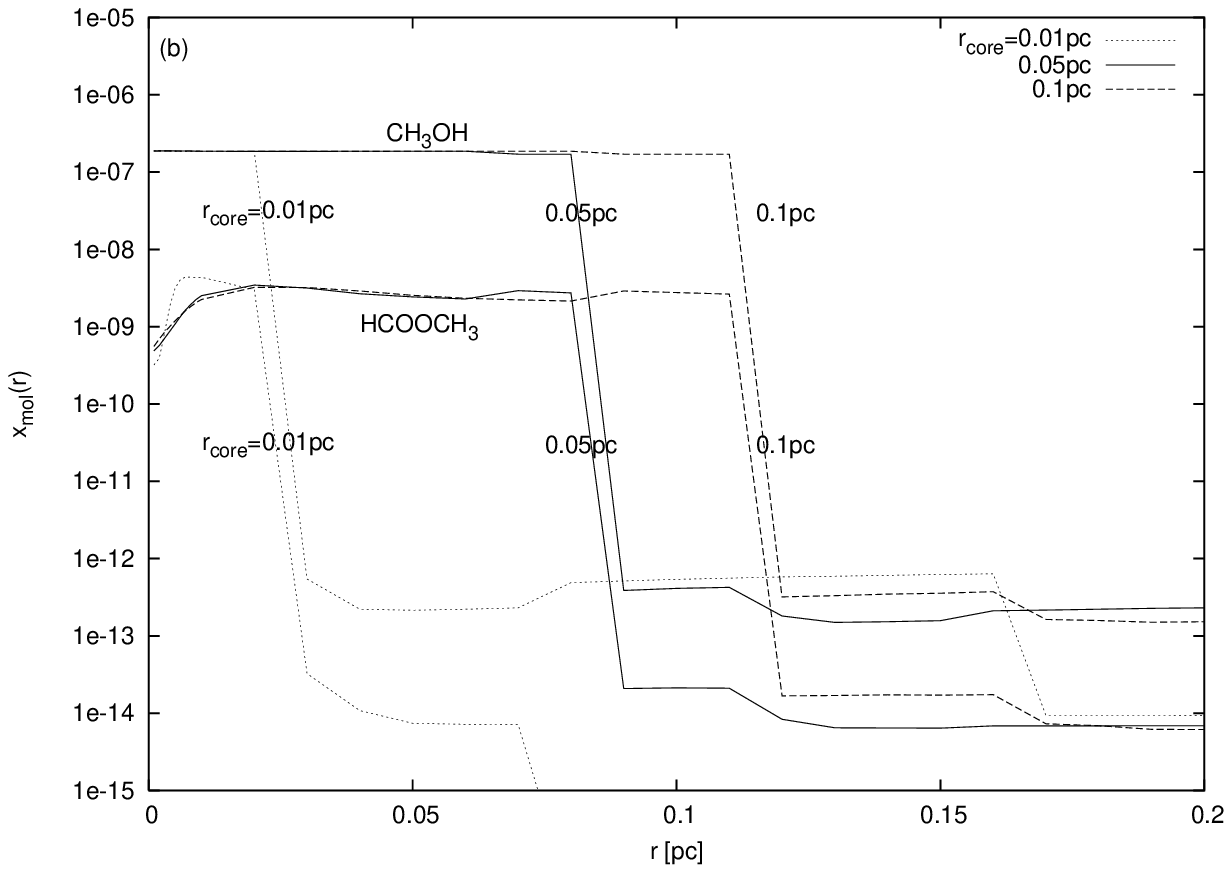}}
\end{figure}
\begin{figure}
\resizebox{\hsize}{!}{\includegraphics{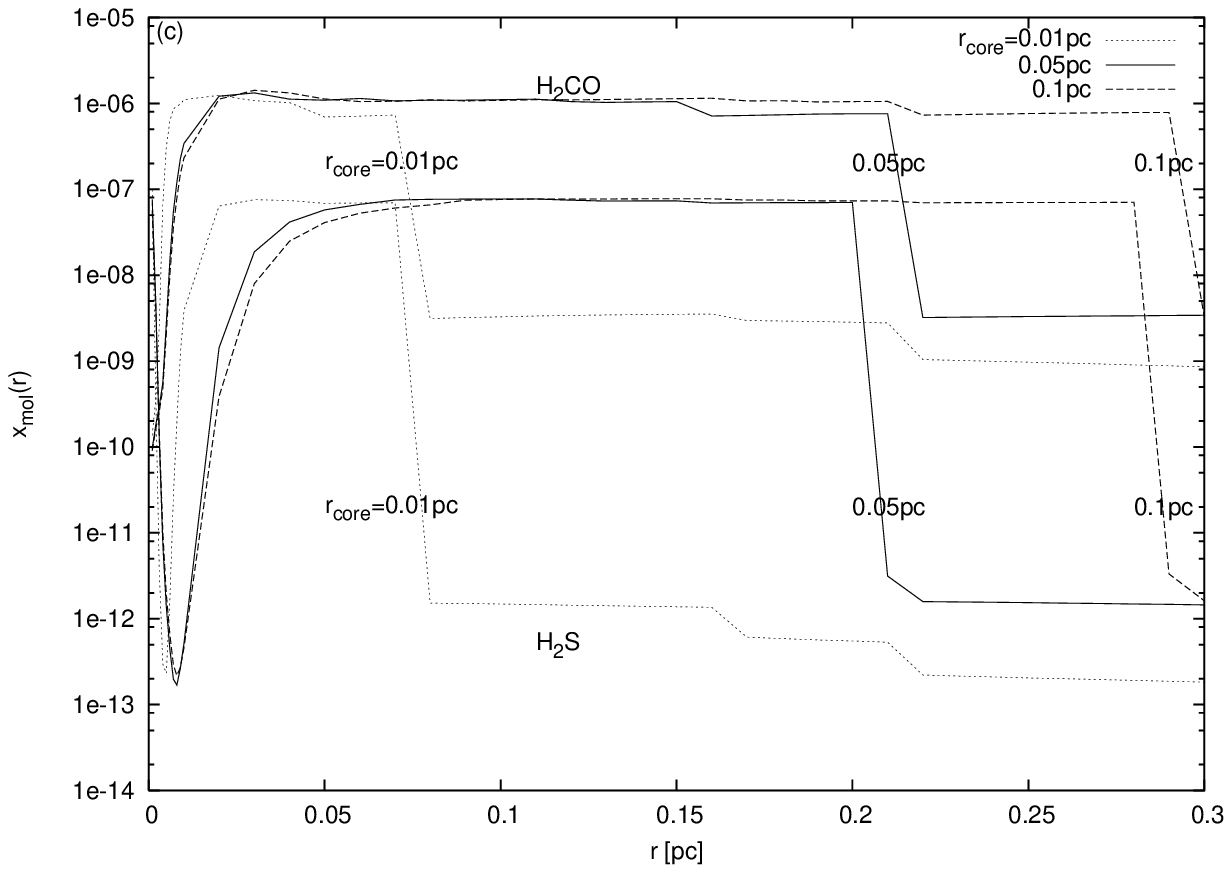}}
\resizebox{\hsize}{!}{\includegraphics{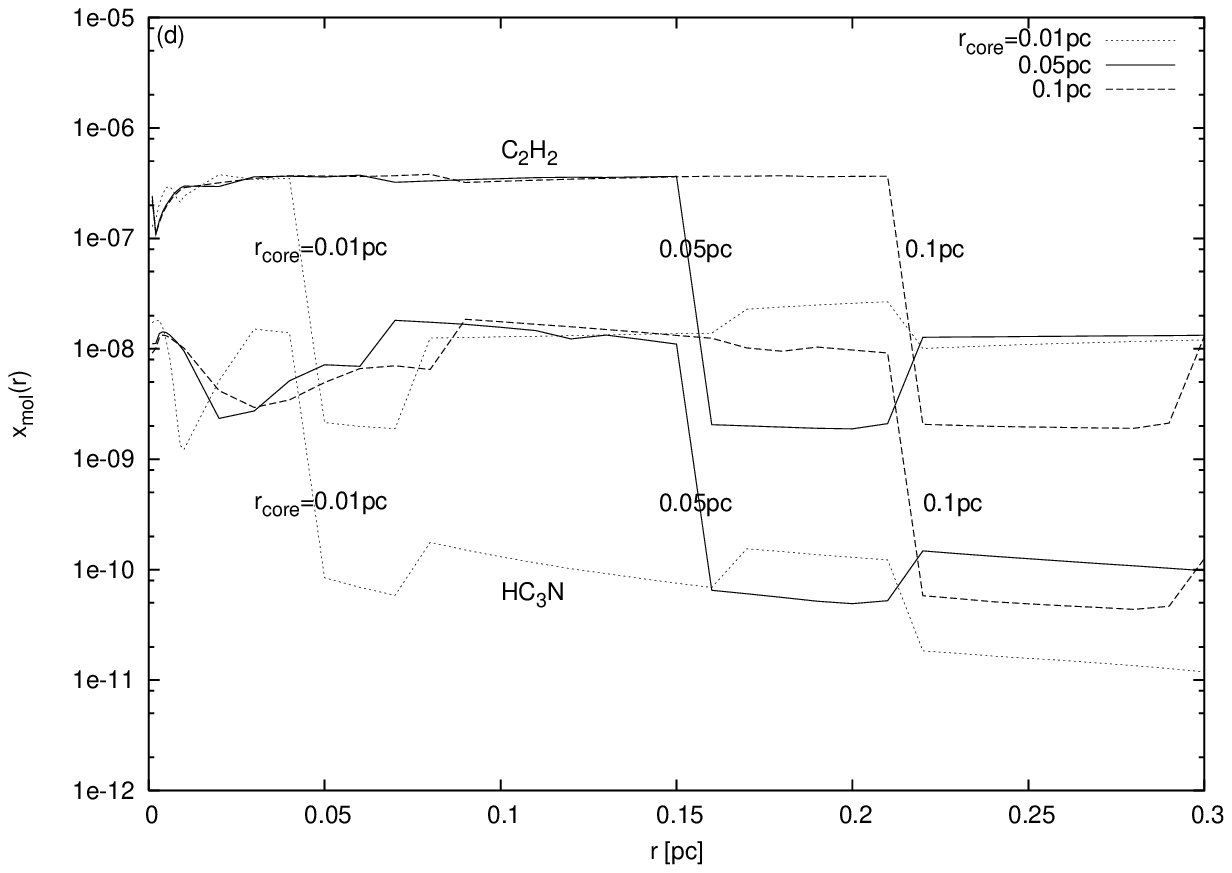}}
\resizebox{\hsize}{!}{\includegraphics{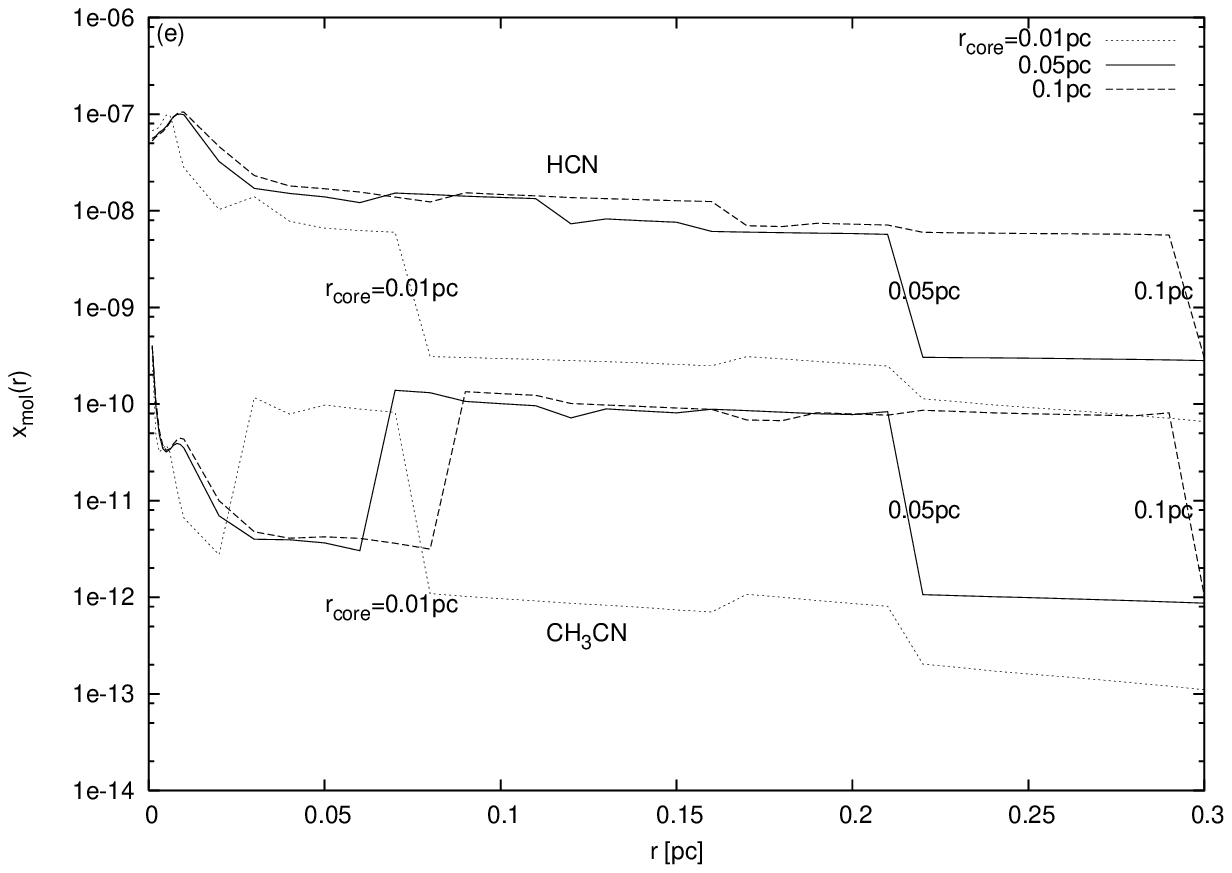}}
\caption{The radial abundance profiles of some parent and daughter
 molecules of the models in the linear radial scales
 with $L_*=10^5L_{\odot}$ and $r_{\rm c}=$
 0.01 (dotted lines), 0.05 (solid lines), and 0.1 pc (dashed lines). 
 The abundance jumps appear at the inner radii in the models with
 the smaller core radii, that is, having steeper density and temperature
 profiles.}
\end{figure}

We calculate the gas-phase chemical reactions in the clumps with
the above density and temperature profiles and $L_*=10^5L_{\odot}$
by means of the chemical model in Sect. 3. Fig.8 gives the resulting radial
abundance profiles of some molecules at $10^4$ yr in the linear radial
scales (while those in Fig.4 are scaled in the logarithmic manner). They
are parent or daughter species and have dramatic changes at certain
radii related to grain mantle evaporation.
H$_2$O, CO$_2$, H$_2$CO, H$_2$S, C$_2$H$_2$, and CH$_3$OH are possible
parent molecules, and have abundance jumps at their injection radii. The
daughter species HC$_3$N, CH$_3$CN, and HCOOCH$_3$ are produced mainly from
C$_2$H$_2$, HCN, and CH$_3$OH, respectively, in this model, and their
abundance jumps appear at the same radii as those of the parent species.
Comparison between reasonably high resolution observations of
abundance jumps and precise radiative transfer and physical-chemical
models of hot cores may make it possible to test the formation
processes of these molecules.

From Fig.8 we can see that the abundance jump of each molecule occurs
at a smaller radius in the clump with a smaller core radius $r_{\rm
c}$, that is, with a steeper density and therefore a steeper
temperature profile. The molecules in Fig.8 can keep high
fractional abundances only
in the inner region of the clump ($< 0.1$pc) if $r_{\rm c}=0.01$pc.
This result suggests that the chemically rich hot cores, whose
observational beam-averaged molecular column densities are large, have
flat density profiles rather than steep ones. On the basis of the low
and high mass star formation theories, the central density profiles of
clumps are thought to be steep just after the star formation, at least if 
accretion of the surrounding material to the central protostellar
cores is important (e.g., Shu 1977; Yorke \& Bodenheimer 1999;
Masunaga \& Inutsuka 2000; Yorke \& Sonnhalter 2002). 
Subsequently, dissipation by outflows will let the clumps
gravitationally unbound and their central density profiles will be 
flattened (e.g., Shu et al. 1987; Nakano et al. 1995). In the case of
massive star formation, powerful outflows with high mass loss rates
as well as radiation pressure and strong winds from the stars may
accelerate the dissipation process. If massive stars are formed as a
result of coalescence of lower mass stars, the envelopes will have
different density profiles from those of isolated star-forming clumps
(e.g., Bonnell et al. 1998).
Statistical observations of various kinds of molecular lines towards hot
cores could constrain their density profiles and the massive star
formation theories.

The effects of inflow motion of hot cores have not been considered in
this paper. If the inflow has a velocity of a few km s$^{-1}$ as is estimated
from the observed atomic and molecular lines (e.g., Keto 2002), it will
not affect the structure on the scale of about 0.25 pc, which
corresponds to the observational beam size for most of the molecular
lines we have used in this paper, within the timescale of $10^4$
yr. Comparison between  model calculations and the forthcoming high spatial
resolution and high sensitivity observations could reveal the dependence
of the chemical structure of hot cores on the inflow motion. In
addition, the temporal evolution of young massive stars in hot cores is also
a factor that may possibly influence the fractional
abundances of some species (cf. Viti \& Williams 1999).

\section{Summary}

We have constructed the density and temperature profiles of hot cores
that have embedded luminous objects, using radiative transfer
calculations together with observations of the spectral energy
distribution. We have used these to investigate hot core chemistry
 taking into account the temperature 
dependent grain mantle evaporation process.

As a result of applying the model to the hot core G34.3+0.15 and
calculating the time-dependent gas-phase reactions in hot gas heated by
a central massive star, it is confirmed that the age of around
$10^4$ yr is preferred to reproduce the large column densities
of both the parent and the daughter molecules. In fact, the calculated
column densities are consistent with most of those observed toward
G34.3+0.15 around 10$^4$ years after central luminous star formation
with a particular initial composition for the grain mantle molecules.

Also, we find that dramatic changes appear in the radial profiles of
fractional abundances of some parent, daughter and related species
at the parents' injection radii, inside which the evaporation time of the
molecules from dust grains are shorter than their accretion time.
The abundance jumps appear because (1) most of the parents are difficult
to destroy within the timescale of 10$^4$ yr, and (2) the initial
abundances of parents is large enough, so that despite their slow 
destruction rates inside the injection radii, they produce appreciable 
abundances of daughter products, while (3) these parent and daughter 
molecules are hardly produced by the gas-phase reactions outside the 
injection radii.

Finally, we discussed the influence of some chemical and physical
properties on the chemical structure of hot cores. First, we investigated
the effects of the trapping of mantle molecules in water ice. As a result,
we find large changes in the location of jumps in the
radial abundance profiles of some species and that it changes 
completely the radial abundance profiles of daughter molecules whose 
production are suppressed in regions where H$_2$O is evaporated 
from grain mantle.
Also, we investigated the chemical structure of hot cores with
different density profiles. Our results suggest that the chemically rich
hot cores have flat density profiles rather than sharp ones. This is
because a steeper radial density profile leads to a steeper
temperature distribution as long as the clump is optically thick to
radiation from the central star.  This means that the abundance jumps
appears at smaller radii and that therefore beam-averaged molecular column
density cannot be large. Observational possibilities of constraining 
the density profiles of hot cores and the massive star formation theories 
are also suggested.

\begin{acknowledgements}
We would like to thank the referee, Dr. F.L. Schoier, for his comments 
which improved our paper.
Astrophysics at UMIST is supported by a grant from PPARC.
\end{acknowledgements}

\end{document}